\documentclass[aps,pra,amsmath,notitlepage,nofootinbib,twocolumn,superscriptaddress,noeprint]{revtex4-2}
\usepackage[utf8]{inputenc}
\usepackage[english]{babel}
\usepackage[T1]{fontenc}
\usepackage{amsmath,amsfonts,amssymb,amsthm,bm,bbm,bbold}
\usepackage{graphicx}
\graphicspath{{Figures/}{./}}
\usepackage{amssymb}
\usepackage{mathtools}
\usepackage{physics}
\usepackage{makecell}
\usepackage{textcomp}
\usepackage{microtype}
\usepackage[dvipsnames]{xcolor}
\usepackage{dsfont}
\usepackage[breaklinks=true,colorlinks=true,linkcolor=teal,urlcolor=teal,citecolor=teal]{hyperref}
\usepackage[caption=false]{subfig}

\newcommand{\I}{\mathrm{i}}
\newcommand{\E}{\mathrm{e}}

\newcommand{\Gperp}{\Gamma_{\!\!\perp}}
\newcommand{\sigplus}{\sigma_{\!+}}
\newcommand{\sigmin}{\sigma_{\!-}}

\renewcommand{\vec}{\boldsymbol}
\let\originalleft\left
\let\originalright\right
\renewcommand{\left}{\mathopen{}\mathclose\bgroup\originalleft}
\renewcommand{\right}{\aftergroup\egroup\originalright}
\renewcommand{\right}{\aftergroup\egroup\originalright}

\usepackage{stackengine}

\begin{document}

\title{On the role of chirping in pulsed single photon spectroscopy}

\author{Elnaz Darsheshdar}
\email{darsheshdare@gmail.com}
\affiliation{Department of Physics, University of Warwick, Coventry, CV4 7AL, United Kingdom}

\author{Aiman Khan}
\affiliation{Department of Physics, University of Warwick, Coventry, CV4 7AL, United Kingdom}
\affiliation{Department of Physics and Astronomy, University of Exeter, Exeter, EX4 4QL, United Kingdom}

\author{Francesco Albarelli}
\affiliation{Scuola Normale Superiore, I-56126 Pisa, Italy}

\author{Animesh Datta}
\email{animesh.datta@warwick.ac.uk}
\affiliation{Department of Physics, University of Warwick, Coventry, CV4 7AL, United Kingdom}


\begin{abstract}
We investigate the precision of estimating the interaction strength between a two-level system (TLS) and a single-photon pulse when the latter is subject to chirping. 
We consider linear, quadratic, and sinusoidal temporal phases applied to Gaussian and exponential temporal profiles. 
At the asymptotic time, when the TLS has fully decayed to its ground state, the fundamental precision depends solely on the magnitude of its spectral amplitude.
For quadratically phase-modulated Gaussian pulses, this is entirely determined by the spectral bandwidth.
We provide expressions for evaluating the fundamental precision for general temporal profiles and phase modulations.
Finally, we show that experimentally feasible mode-resolved measurements are optimal, or close to it, for chirped, pulsed single photon spectroscopy.
\end{abstract}

\maketitle

\section{Introduction}

Chirped pulses of laser light -- that is those with time-dependent frequency have a long history in optimal control of quantum matter systems \cite{Amstrup1993,Kohler1995} and their spectroscopy \cite{Garraway1995,Polli2010}. 
Presently, efforts are underway to investigate and identify the potential advantages of using quantum states of light in spectroscopy~\cite{Mukamel2020} with the aim of surpassing the performance of pulses of classical states of light, namely coherent states~\cite{kalachev2007biphoton,kalashnikov2014,Dorfman2021b}.

In a general sense, quantum light spectroscopy studies the problem of extracting information about a quantum matter system indirectly by measuring the quantum state of the light pulse that has interacted with it.
A meaningful theoretical analysis of the performances of quantum pulses of light for the task of estimating parameters of the matter system requires looking at the problem with the lens of quantum estimation theory~\cite{albarelli2022fundamental,khan2023does}.

Experimentally, the ability to sculpt the spectral-temporal waveform of quantum states of light is improving~\cite{Karpi_ski_2016,So_nicki_2023}. 
Indeed, the control and manipulation of quantum light pulses for quantum information processing~\cite{Karpinski2021} is advancing more widely.
These motivate the exploration of the role of time-dependent phase or frequency in pulsed quantum light spectroscopy.

In this paper, we investigate how a single-photon pulse with a temporal profile of
\begin{equation}
\label{eq:pulse}
\xi(t) = \xi_R(t)e^{i\phi(t)},~~\xi_R(t),\phi(t) : \mathbb{R} \to \mathbb{R}.
\end{equation} 
performs in estimating $\Gamma$ -- the coupling constant of a two-level system (TLS) with the photon.
For this simplest of matter systems, we explore the consequences of linear, quadratic, and sinusoidal $\phi(t)$ on
the fundamental and attainable limits of the precision of estimating $\Gamma.$
This paper thus goes beyond Ref.~\cite{albarelli2022fundamental} where only real-valued pulses with $\phi(t)=0$ were studied.

Our main results are as follows:
\begin{enumerate}
\item At the asymptotic time, the quantum Fisher information (QFI) for estimating $\Gamma $ depends on the probability distribution of the single-photon wave-packet in the frequency domain and the central frequency of the pulse. See Eqs.~\eqref{eq:QFI_1ph_normalized} and \eqref{eq:pgamma}--\eqref{eq:qfigamma2}.

\item For a quadratic temporal phase modulation applied to a real Gaussian pulse, the asymptotic QFI only depends on the initial input wave-packet through its frequency bandwidth (assuming the same central frequency). See Sec.~\ref{sec:quad}.
However, this does not hold more generally.

\item For finite times, we provide a general expression for the QFI which can be examined to determine the benefits of $\phi(t)$ in estimating
 $\Gamma$. See Appendix~\ref{App:A}.

\item For phase-modulated pulses symmetric in frequency, we analytically construct (temporal) mode-resolved measurements that saturate the fundamental limits.
These are also near-optimal for other modulation scenarios, e.g. linear phase modulation. See Sec. \ref{sub:overlapzero}.
\end{enumerate}

The paper is structured as follows. Sec.~\ref{sec:theory} presents our theoretical framework, encompassing the model of light-matter interaction and a summary of local quantum estimation theory.
Sec.~\ref{sec:pmp} contains new expressions for the QFI in the frequency domain, which are instrumental for studying the impact of phase modulation. Sec.~\ref{sec:phase} provides a comprehensive analysis of the precision limits associated with estimating $\Gamma$ using temporally phase-modulated single-photon pulses.
We focus on scenarios where the phase modulates a Gaussian real-valued pulse, exploring the impact of linear, quadratic, and sinusoidal phase modulations across various TLS-environment (T-E) coupling values. Sec.~\ref{exponential} explores the effect of linear and quadratic phase modulation applied on exponentially decaying real-valued single-photon pulses.
Sec.\ref{sec:Measurement} presents an investigation into optimal and near-optimal measurements achieved through mode-resolved photon counting.
Sec.~\ref{sec:Conclusions} concludes with a concise summary and discussion.

\begin{figure*}[th]
    \centering
    \includegraphics[width=0.9\textwidth]{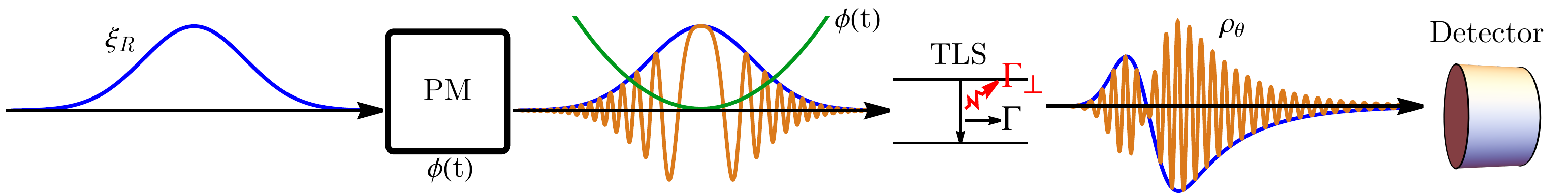}
    \caption{Illustration (not to scale) of a complex-valued pulse $\xi(t) = \xi_R(t)e^{i\phi(t)}$ interacting with the TLS. $\Gamma$ ($\Gamma_\perp$) captures the interaction strength with the pulse (environment).
     A phase modulator (PM) applies a modulation $e^{i\phi(t)}$ to a real pulse $\xi_R(t)$. For instance, $\phi(t)=K{{t}^{2}}=k{{\Gamma}^{2}}{{t}^{2}}$ which is a quadratic chirping is represented by the green curve.
     $\rho_{\theta}$ is given by Eq.~\eqref{eq:outgoing_state}.}
    \label{fig:setup}
\end{figure*}

\section{Theoretical framework} \label{sec:theory}

\subsection{TLS, Light and Environment Interaction}

We consider a TLS, denoted by the $\mathrm{T}$ subsystem, fixed in space and described by the free Hamiltonian, 
\begin{equation}
H^{\mathrm{T}} = \hbar \omega_0 \dyad{e}{e},
\end{equation}
where the ground and excited states of the TLS are denoted by $\ket{g}$ and $\ket{e}$ respectively, with the ground state energy set to zero, and  $\omega_0$ the transition frequency. 

The light is a traveling pulse of the quantized radiation field, denoted by the $\mathrm{P}$ subsystem, described by a continuum of frequencies and has a well-defined direction of propagation with the free field Hamiltonian~\cite{Blow1990a}
 \begin{equation}
    \label{eq:HF}
    H^\mathrm{P} = \hbar \int_0^\infty \! d\omega \, \omega \, a^\dag (\omega) a(\omega),
 \end{equation}
where the bosonic operators are labeled by a continuous frequency $\omega$ and satisfy the commutation relation $[a(\omega),a^\dag(\omega')]=\delta(\omega-\omega')$. The electric field operator, corresponding to a well-defined direction of propagation, is 
\begin{equation}\label{eq:Electric_field_1D_continuous}
\mathbf{E}(t) = \I \hbar \bm{\epsilon} \mathcal{A}(\bar{\omega})  \int_{-\infty}^{\infty} \frac{d\omega}{\sqrt{2\pi}} \; a(\omega) \E^{-i \omega t},
\end{equation}
where $\bm{\epsilon}$ represents the unit polarization vector. The function $\mathcal{A}(\bar{\omega})$ is defined as $\sqrt{\frac{\bar{\omega}}{2 \epsilon_{0}c A \hbar}}$, where $\bar{\omega}$ denotes the central frequency of the pulse, and $A$ refers to the transverse quantization area.
The integral in Eq.(\ref{eq:Electric_field_1D_continuous}) spans the entire $\omega$-axis, incorporating the slowly-varying envelope approximation\cite{Schlawin2017,PhysRevA.43.2498} that is applicable to paraxial pulses considered here.

A schematic depiction of the interaction between the traveling pulse and the TLS is shown in Fig. \ref{fig:setup}.
As the matter system (TLS) is much smaller in scale than optical wavelengths, we invoke the dipole approximation~\cite{mandel1995optical} wherein the TLS-light interaction takes the form $-\vec{\mu} \cdot \vec{E}$, where $\vec{\mu}$ is the transition dipole moment operator, and $\vec{E}$ is the quantized electric field. 
In the picture generated by the unitary transformation $e^{-\I H_0t}$, where~\cite{khan2023does}
\begin{equation}\label{eq:zerothHamiltonian}
    H_0 = \hbar \bar{\omega} \dyad{e}{e} + \int_0^{\infty} d\omega~\hbar\omega~a^{\dag}(\omega)a(\omega),
\end{equation}
and using the slowly-varying envelope \cite{Ko2022} and the rotating wave approximations,
the TLS-pulse (TP) Hamiltonian is given by~\cite{Scully1997,Ko2022}
\begin{align}
\label{eq:interaction_hamiltonian_initial_definition}
H^{\mathrm{TP}}(t)  = \hbar \Delta \dyad{e}{e} -\I \hbar \sqrt{\Gamma} \left( \sigma_{\!+} a(t) - \sigma_{\!-} a^\dag(t)  \right),
\end{align}
where $\Delta=\omega_0-\bar{\omega}$ and $\sigmin = \dyad{g}{e}=\sigplus^{\dagger}$ is the TLS dipole operator. 
In Eq. \eqref{eq:interaction_hamiltonian_initial_definition} we have introduced the so-called ``quantum white-noise'' operators \cite{Ko2022}),
\begin{equation}
    \label{eq:white_noise_a}
    a(t) = \int_{-\infty}^{\infty} \frac{d\omega}{\sqrt{2\pi}} \; a(\omega) \E^{-\I (\omega-\bar{\omega}) t},
\end{equation}
satisfying $[a(t),a^\dag(t')] = \delta(t-t')$ as well as the coupling constant $\Gamma=(\bm{\mu}_{eg}\cdot \bm{\epsilon})^2 \mathcal{A}(\bar{\omega})^2$ which is proportional to the square of the dipole moment.

In addition to the continuum of bosonic modes $a(\omega)$ that describe the pulse degrees of freedom, the TLS in free space interacts with an infinitude of other modes of the electromagnetic field, capturing all the other spatial and polarization degrees of freedom beyond those of the pulse.
Description of the infinitely large electromagnetic environment is drastically simplified by the fact that it is inaccessible to experiments, 
and can be considered in terms of its effects on the reduced dynamics. We account for this by introducing a coupling to an additional continuum of bosonic modes $b(\omega)$ with a free Hamiltonian of $H^\mathrm{E} = \hbar \int_0^\infty \! d\omega \, \omega \, b^\dag (\omega) b(\omega)$, leading to the final form for the TLS-pulse-environment (TPE) Hamiltonian
\begin{equation}
    \label{eq:Hint_PE}
	H^\mathrm{TPE}(t) =  \hbar \Delta \dyad{e}{e} -\I \hbar  \left( \sqrt{\Gamma} a(t)\sigma_{\!+} + \sqrt{\Gperp } b(t)\sigma_{\!+}- \mathrm{h.c.} \right) ,
\end{equation}
where the additional set of white noise operators $b(t)$ satisfying $[b(t),b^\dag(t')] = \delta(t-t')$ represent an electromagnetic ``environment'' ($\mathrm{E}$) subsystem coupled to the TLS. The coupling strength with this environment is denoted by $\Gamma_\perp$. This representation is analogous to a more realistic model where the environment is composed of a discrete set of infinitely many families of white noise operators, accounting for all degrees of the electromagnetic field beyond those described by $a(\omega)$~\cite[Appendix A]{albarelli2022fundamental}.

The Hamiltonian in Eq.~\eqref{eq:Hint_PE} can describe various light-matter-environment coupling geometries, simply by assigning different relative magnitudes for the quantities $\Gamma_\perp$ and $\Gamma$. The relative magnitudes can be manipulated depending on different experimental setups \cite{Silberfarb2004,Wang2011,Molmer2024}. In the following, we refer to the choice
 $\Gamma_\perp=0$ as perfect TLS-pulse (T-P) coupling, while $\Gamma_\perp=5\Gamma$ as strong T-E coupling, 
 with the idea of capturing roughly a free space geometry.
We treat T, P, and E as distinct subsystems, with P being the only one that can be experimentally measured, while the light emitted into the environment E is irreversibly lost. 

\subsection{Unitary evolution of the TLS, pulse and environment}
\label{subsubsec:1photonEvo}

We focus on a single-photon pulse 
\begin{equation}
\ket{1_{\xi}} = \int_{-\infty}^\infty d\tau ~\xi(\tau) a^\dag (\tau)  \ket{0^\mathrm{P}},
\end{equation}
where $\xi(\tau)$ is a complex function of time.
Assuming that the TLS starts in the ground state $\ket{g}$ and the environment in vacuum $ \ket{0^\mathrm{E}},$
the joint TPE quantum state after evolution
under the Hamiltonian in Eq.~\eqref{eq:Hint_PE} is~\cite{albarelli2022fundamental}
\begin{equation}
    \label{eq:1ph_pulse_full_state}
    \ket{\Psi^\mathrm{TPE}} = \psi_e \ket{e} \ket{0^\mathrm{P}} \ket{0^\mathrm{E}} + \ket{g} \left( \ket*{\widetilde{\psi}_{g}^\mathrm{P}} \ket{0^{\mathrm{E}}} + \ket{0^\mathrm{P}} \ket*{ \widetilde{\psi}_{g}^\mathrm{E} } \right),
\end{equation}
where $ \psi_e$ is the excitation amplitude of the TLS, and
\begin{eqnarray}
\ket*{\widetilde{\psi}_{g}^\mathrm{P}(t)} &=& \int_{-\infty}^\infty d\tau \widetilde{\psi}_g^\mathrm{P}(t,\tau) a^\dag (\tau)  \ket{0^\mathrm{P}},  \\
\ket{\widetilde{\psi}_g^\mathrm{E}(t)} &=& \int_{-\infty}^\infty d\tau \widetilde{\psi}_{g}^\mathrm{E}(t,\tau) b^\dag (\tau)  \ket{0^\mathrm{E}},
\end{eqnarray}
represent unnormalized single-photon states in the pulse and environment modes, respectively. 
More explicitly,~\cite{Konyk2016,Ko2022}
\begin{align}\label{eq:1ph_pulse_psie} 
    \psi_e(t) &= - \sqrt{ \Gamma} \int_{t_0}^t \! dt' \, e^{-((\Gamma + \Gamma_\perp)/{2}+\I \Delta)(t-t')} \xi(t'),  \\ 
    \label{eq:1ph_pulse_unnorm_state}
    \ket*{\widetilde{\psi}_{g}^\mathrm{P}(t)} &= \int_{t_0}^\infty \! d\tau \left( \xi(\tau) + \sqrt{\Gamma} \Theta(t-\tau) \psi_e(\tau) \right) a^\dag(\tau) \ket{0^\mathrm{P}}, \\
    \label{eq:1ph_env_state}
    \ket*{\widetilde{\psi}_{g}^\mathrm{E}(t)} &= \sqrt{\Gamma_\perp} \int_{t_0}^t \! d\tau \, \psi_e(\tau) b^\dag (\tau) \ket{0^\mathrm{E}}, 
\end{align}
where $\Theta(x)$ is the Heaviside step function. 
These show that the matter-dependent distortion of the scattered pulse depends on $\psi_e(t)$, which is the convolution of two functions -- pulse shape $\xi(t)$, and the matter characteristic function $\sqrt{\Gamma}\mathrm{exp}\left[-(({\Gamma+\Gamma_{\perp}})/{2}+i\Delta)t \right]\Theta(t)$).

To obtain the P subsystem, we trace out both the T and E subsystems.
This leads to an incoherent mixture of the vacuum and a modified single-photon wave-packet 
\begin{align}\label{eq:outgoing_state}
    \rho_{\theta}= p_\theta \dyad{0}{0} + (1-p_\theta )\dyad{\psi_\theta}{\psi_\theta},
\end{align}
where $\theta$ denotes the parameter of interest, $p_\theta$ and normalized single-photon state are respectively, 
\begin{equation}
p_\theta =  |\psi_e|^2 + \braket*{ \widetilde{\psi}_{g}^\mathrm{E}}{ \widetilde{\psi}_{g}^\mathrm{E} }, ~~~
\ket{\psi_\theta} = \ket*{\widetilde{\psi}_{g}^\mathrm{P}} / \sqrt{\braket*{\widetilde{\psi}_{g}^\mathrm{P}}{\widetilde{\psi}_{g}^\mathrm{P}}},
\label{16}
\end{equation}
such that $\braket{\psi_\theta}{\psi_\theta} = 1$.
\subsection{Quantum light spectroscopy as estimation}

Spectroscopy of a quantum system with pulsed quantum states of light involves probing it through traveling field states that convey information regarding the matter parameters to the output modes of the light. Mathematically, this corresponds to estimating the parameter $\theta$ from the detectable quantum state of the light $\rho_{\theta}$.

The parametric classical statistical model obtained by performing the positive operator valued measurements (POVM)~$\{M_i:~ M_i>0,\, \sum_i M_i = \mathds{I}^{\mathrm{P}}\}$ on the output state $\rho_{\theta}$ is given by the Born rule $\{p(i|\theta) = \mathrm{Tr}\left[\,\rho_{\theta}M_i\,\right]\,|\,\theta\in\mathds{R}\}$ where $\Theta \subseteq \mathds{R}$ is the parameter space. Statistical inference then involves constructing estimators $\hat{\theta} = \theta(X_1,X_2,\dots,X_n)$, where $\{X_i\}$ are random variables corresponding to each of $n$ observed values, independent and identically distributed. For any unbiased estimator or more generally, in the limit of a large number $n$ of experiment repetitions, the estimation error is lower bounded by the Cram\'{e}r-Rao bound~(CRB)~\cite{rao1992information,cramer2016mathematical},
\begin{equation}
    V(\theta|\{M_i\}) \geq \frac{1}{n\,\mathcal{C}(\theta|\{M_i\})}
\end{equation}
where $V(\theta|\{M_i\}) = \mathbb{E}_{\theta}[(\hat{\theta}-\theta)]^2$ is the mean square error~(MSE) of the estimator ~($\mathbb{E}_{\theta}$ denotes expectation with respect to $X_1,X_2,\dots,X_n~\thicksim~p(i,\theta))$,  and $\mathcal{C}(\theta|\{M_i\})$ is the classical Fisher information (CFI), defined as
\begin{equation}\label{eq:cfidefinition}
    \mathcal{C}(\theta|\{M_i\}) = \mathrm{Var}_{\theta}\left[ \frac{\partial}{\partial \theta} \mathrm{log}~ p(i|\theta) \right] = -\mathbb{E}_{\theta}\left[ \frac{\partial}{\partial \theta}\mathrm{log}~ p(i|\theta)  \right]^2.
\end{equation}

The parametric model, and therefore the optimal estimators themselves, depend on the quantum measurement characterized by the POVM $\{M_i\}$.

Minimizing the estimator MSE over all possible quantum measurements yields a new, more fundamental bound for the precision of estimating the parameter $\theta$, given  the quantum parametric model $\{\rho_\theta \,|\,\theta\in \Theta \}$
\begin{equation}
    V(\theta|\{M_i\}) \geq \frac{1}{n\mathcal{C}(\theta|\{M_i\})} \geq \frac{1}{n\mathcal{Q}(\theta;\rho_{\theta})},  
\end{equation}
known as the quantum Cr\'{a}mer-Rao bound~(QCRB)~\cite{holevo2011probabilistic,braunstein1994statistical,paris2009quantum}.
$\mathcal{Q}(\theta;\rho_{\theta})$ is the QFI

\begin{equation}
    \mathcal{Q}(\theta;\rho_{\theta})  = \mathrm{Tr}\,(\,\rho_{\theta}\,L_{\theta}^2\,)  \geq \mathcal{C} (\theta|\{M_i\}).
\end{equation}
defined via the self-adjoint symmetric logarithmic derivative~(SLD) operators, which satisfy the Lyapunov equation
\begin{equation}\label{eq:lyapunovsld}
   L_{\theta}\,\rho_{\theta} + \rho_{\theta}\,L_{\theta} = 2\,\frac{\partial \rho_{\theta}}{\partial \theta}.
\end{equation}

In this paper, we restrict ourselves to single-parameter estimation, for which the corresponding QCRB is necessarily saturated by the projective measurement corresponding to eigenvectors of the SLD operator $L_{\theta}$~\cite{paris2009quantum}. 
For rank-deficient models however, this is only a necessary condition and eigenvectors of the SLD operator are only one of many QCRB-saturating POVMs~\cite{kurdzialek2022measurement}. 

Finally, we introduce two expressions for the QFI that will be employed in later sections.
For a general mixed state with spectral decomposition  ${{\rho }_{\theta }}=\sum\limits_{n}{}{{p}_{n}}\left| {{\psi }_{n}} \right\rangle \left\langle  {{\psi }_{n}} \right|$, the QFI is~\cite{liu2014quantum,liu2019quantum}
\begin{align}\label{eq:qfiarbitraryrank}
&\mathcal{Q}\left(\theta;\sum_{n}\,p_n\,\ket{\psi_{n}}\bra{\psi_{n}}\right)  = \sum_{n}\,\frac{(\partial_{\theta} p_n)^2}{p_n}\noindent\\
    &+ \sum_n\,4p_n\,\langle\partial_{\theta}\psi_{n}|\partial_{\theta}\psi_{n}\rangle
    -\sum_{m,n}\frac{8p_m p_n}{p_m + p_n}\,|\langle\partial_{\theta}\psi_{m}|\psi_{n}\rangle|^2 . \nonumber
\end{align}

Thus, for pure states $\rho_{\theta} = \ket{\psi_{\theta}}\bra{\psi_{\theta}}$, the QFI is
\begin{equation}\label{eq:qfipure}
    \mathcal{Q}(\ket{\psi_\theta}) = 4\,\left(\,\langle\partial_{\theta}\psi_{\theta}|\partial_{\theta}\psi_{\theta}\rangle - |\langle\psi_{\theta}|\partial_{\theta}\psi_{\theta}\rangle|^2\,\right).
\end{equation}

As our $H_0$ in Eq.~\eqref{eq:zerothHamiltonian} does not explicitly depend on the matter parameters that we seek to estimate, the QFI obtained from the quantum state in the interaction picture matches that from the Schr\"{o}dinger. 

\subsubsection{Single-photon QFI: classical and quantum contributions}

From Eq. (\ref{eq:qfiarbitraryrank}), the QFI of $\rho_{\theta}$ in Eq.~\eqref{eq:outgoing_state} is the sum of CFI of the two-outcome probability distribution $\{ p_\theta,1-p_\theta \} $ 
and the QFI of the modified pure single-photon state $\ket{\psi_\theta}$, rescaled by the corresponding probability \cite{albarelli2022fundamental,liu2014quantum}
\begin{align}\label{eq:QFI_1ph_normalized}
    \mathcal{Q}(\rho_{\theta}) &= \frac{ (\partial_\theta p_\theta)^2 }{p_\theta (1-p_\theta)} + (1-p_\theta) \mathcal{Q}\left( \ket{\psi_\theta} \right) \\ \nonumber
    & \equiv ~ \mathcal{C}(p_\theta) + \tilde{\mathcal{Q}}(\ket{\psi_\theta}),
\end{align}
which can be interpreted as classical and quantum contributions respectively~\cite[Sec. III.A.2]{albarelli2022fundamental}.

The QFI is a dimensional quantity when the parameter possesses physical dimensions. To facilitate comparisons across various parameter values, our emphasis will be on the dimensionless QFI, denoted as $\theta^2 \mathcal{Q}(\rho_{\theta})$, which signifies the inverse of estimation precision relative to the true parameter value.

\section{Chirped pulses}
\label{sec:pmp}

We now present new results on the role of a temporal phase on the fundamental and attainable precision of estimating the parameter $\Gamma$ 
for a single-photon pulse of the form in Eq.~\eqref{eq:pulse}.
An immediate consequence is the presence of the term $\left\langle  {{\psi }_{\theta }} | {{\partial }_{\theta }}{{\psi }_{\theta }} \right\rangle$ in Eq.~\eqref{eq:qfipure} which is 
nonzero\footnote{It is zero for $\phi(t) = 0$ and for complex-valued pulses that are symmetric in the frequency domain. See Appendix \ref{app:realvec_pulses}.} for $\phi(t) \neq0.$
While this may naively suggest a reduction of the QFI, it is not necessarily so as the first term also changes for $\phi(t) \neq0.$ See Sec. \ref{exponential} for an instance and Sec. \ref{sec:Measurement} for its effect on the CFI.

General expressions for evaluating the two contributions in Eq. (\ref{eq:QFI_1ph_normalized}) for $\theta \equiv \Gamma$ at finite times are provided in Appendix \ref{App:A}. 
The time domain expressions therein seem more amenable for finite-time calculations.
Frequency domain expressions, however, seem more amenable for calculations of the quantities at asymptotic time. 
They are presented next.


At the asymptotic time $t \gg 1/\mathrm{max}(\Gamma,\Gamma_{\perp})$, the TLS decays to the ground state and becomes disentangled from the light.
However, for $\Gamma_\perp > 0$, the state of the light is not pure, since the photon of the pulse is partly lost to the environment. 

We first express the unnormalized single-photon states in the P and E modes (Eqs. (\ref{eq:1ph_pulse_unnorm_state}) and (\ref{eq:1ph_env_state})) in the frequency domain. 
Using the convolution theorem, these become
\begin{align}
    \left| {\tilde{\psi}_{g}^{P}}(\infty ) \right\rangle =\int_{-\infty }^{\infty }{d}\omega \tilde{\xi} (\omega ) \left( 1-\sqrt{ \Gamma}f(\omega) \right){{a}^{\dagger }}(\omega )\left| {{0}^{P}} \right\rangle,
\end{align}
\begin{align}
    \left| {\tilde{\psi}_{g}^{E}}(\infty ) \right\rangle =-\sqrt{ \Gamma_\perp} \int_{-\infty }^{\infty }{d}\omega \tilde{\xi} (\omega )f(\omega) {{b}^{\dagger }}(\omega )\left| {{0}^{E}} \right\rangle,
\end{align}
where 
\begin{equation}
\label{eq:char}
f(\omega)= \frac{\sqrt{{\Gamma} }}{ \left({\Gamma +\Gamma_\perp}\right)/{2}-\I(\omega-\Delta)  },
\end{equation}
is the frequency domain characteristic function of the TLS. 
See Appendix \ref{App:B} for finite-time expressions in the frequency domain.

\begin{figure*}[ht]
    \subfloat[ \label{fig:1anew}]{
	\includegraphics[width=0.28\textwidth]{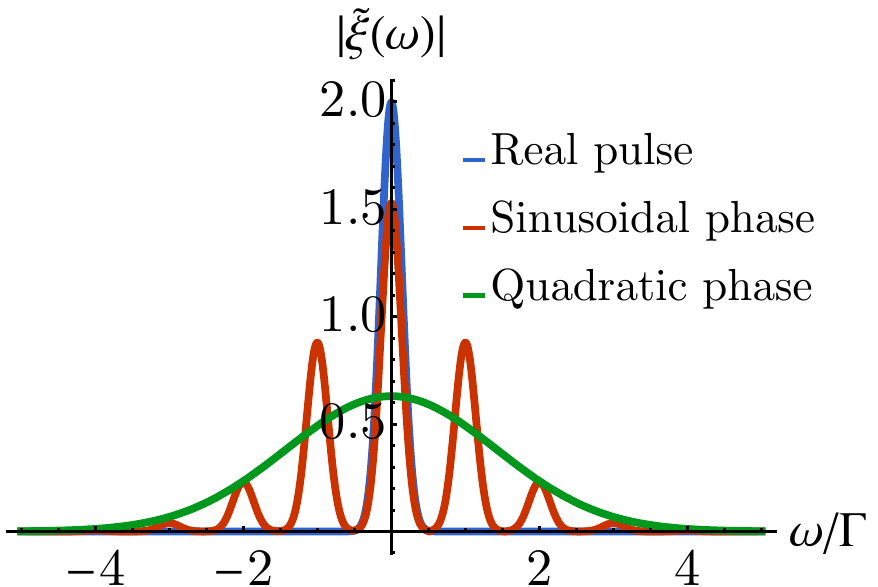}
	}
    \subfloat[\label{fig:1bnew}]{
	\includegraphics[width=0.28\textwidth]{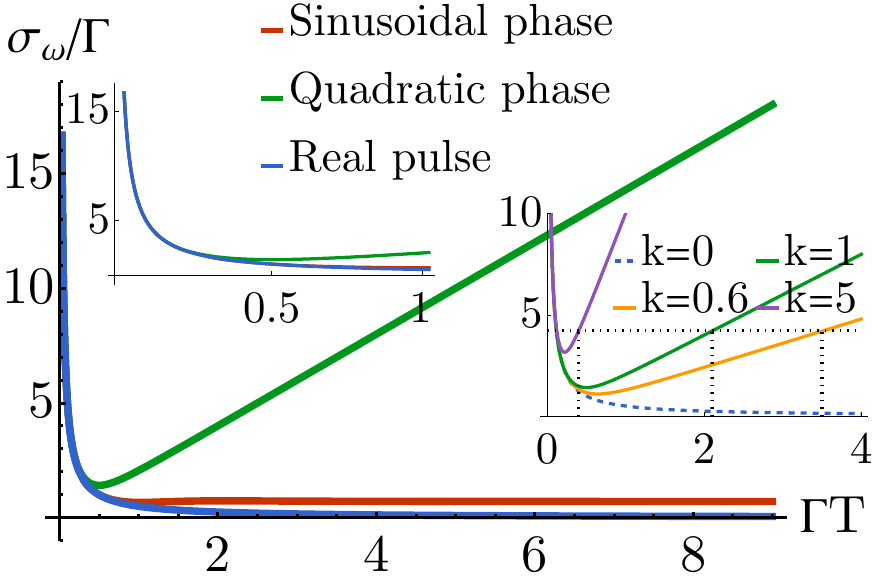}
	}
    \subfloat[ \label{fig:1cnew}]{
	\includegraphics[width=0.28\textwidth]{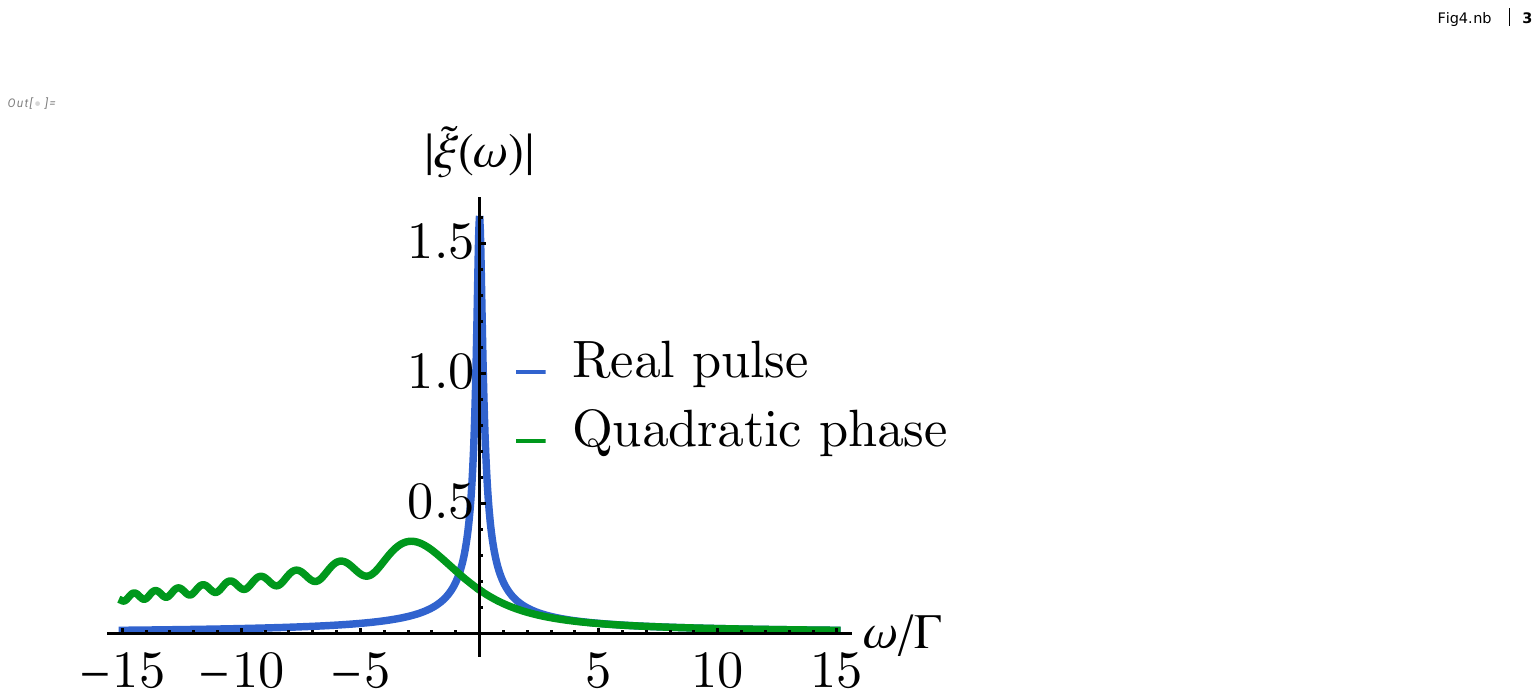}
	}
    \caption{ (a) A real Gaussian pulse along with its quadratic and sinusoidal phase-modulated counterparts using parameters $\Omega=1$, and $k=0.1$. 
    		  (b) The frequency domain bandwidths of each pulse featured in (a) against their respective time durations using parameters $\Omega=1, k=1$. The area where the three lines are closely clustered is depicted in the upper inset. The second (bottom) inset illustrates the changes in bandwidth for various values of $k$. (c) A real exponentially decaying pulse along with its quadratic phase-modulated counterpart using parameter $k=1$. }
    \label{fig:1new}
\end{figure*}

The asymptotic expressions of $p_\theta$ and its derivative in Eq. (\ref{eq:QFI_1ph_normalized}) for $\theta \equiv \Gamma$ are 
\begin{align}\label{eq:pgamma}
    {{p}_{\Gamma }}(\infty) 
    =\Gamma_\perp\int_{-\infty }^{\infty }{d}\omega {\left| \tilde{\xi} (\omega) f(\omega) \right|}^{2}
\end{align}
and
\begin{align}\label{eq:dpgamma}
    {{\partial }_{\Gamma }}{{p}_{\Gamma }}\left( \infty  \right)={\Gamma_\perp}\int_{-\infty }^{\infty }{d}\omega {{\left| \tilde{\xi} (\omega) \right|}^{2}} \partial_\Gamma \left|f(\omega)\right|^{2}
\end{align}
respectively.
The two terms of the quantum contribution in Eq. (\ref{eq:qfipure}),
in the frequency domain are 
\begin{align}\label{eq:qfigamma1}
    &\left\langle {{\partial }_{\Gamma }}{\tilde{\psi}_{g}^{P}}(\infty ) \right.\left| {{\partial }_{\Gamma }}{\tilde{\psi}_{g}^{P}}(\infty ) \right\rangle =\\ \nonumber
    &\frac{1}{4\Gamma}\int_{-\infty }^{\infty }{d}\omega {{\left| \tilde{\xi} (\omega ) \right|}^{2}}  \left | f(\omega) + 2\Gamma \partial_\Gamma f(\omega) \right |^2, 
\end{align}
and
\begin{align}\label{eq:qfigamma2}
    &\left\langle {\tilde{\psi}_{g}^{P}}(\infty ) \right.\left| {{\partial }_{\Gamma }}{\tilde{\psi}_{g}^{P}}(\infty ) \right\rangle \\ \nonumber
    = 
    & -\frac{1}{2\sqrt{\Gamma}}\int_{-\infty }^{\infty }{d}\omega {{\left| \tilde{\xi} (\omega ) \right|}^{2}}\left( 1-\sqrt{ \Gamma}f^*(\omega) \right) \left( f(\omega) + 2\Gamma \partial_\Gamma f(\omega) \right)
\end{align}
respectively.

Eqs.~(\ref{eq:pgamma}-\ref{eq:qfigamma2}) show that both the classical and quantum contributions to the asymptotic QFI 
depend only on ${{\left| \tilde{\xi} (\omega ) \right|}^{2}}$ of the incident single-photon pulse.
These expressions can be used to shape pulses for ``optimal quantum light spectroscopy'' -- in line with optimal quantum control. The exploration of optimal pulse shape using this expression will be discussed in another publication \cite{Das2025}.   
This also shows that the spectral phase that precedes the temporal phase in several experiments~\cite{Karpinski2021,torres2011space,Karpi_ski_2016,So_nicki_2023,Yu_2022,Zhu_2022,Bennett_2001,sosnicki2020aperiodic,Karpinski2017} 
has no impact except via a different  $ \xi_R(t).$

We focus here on a few instances of pulse shapes $ \xi_R(t)$ to highlight the potential and rich behaviour of chirping in quantum-light spectroscopy. Our choices, such as a
Gaussian temporal profile and quadratic and sinusoidal temporal phase, are motivated by recent experimental efforts~\cite{Karpi_ski_2016,Karpinski2017,Karpinski2021}.
Fig. \ref{fig:1anew} depicts the frequency domain pulse shapes of a real Gaussian pulse, as well as its counterparts with quadratic and sinusoidal phase modulations (for $k=0.1$ and $\Omega=1$). In Fig. \ref{fig:1bnew}, we illustrate the manipulation of frequency domain bandwidth ($\sigma _{\omega }$) through quadratic and sinusoidal time phase modulations (for $k=1$ and $\Omega=1$). The quadratic phase modulation has a much more pronounced effect on bandwidth alteration compared to the sinusoidal phase.
Note that, due to the significantly greater impact of quadratic phase modulation on pulse bandwidth compared to sinusoidal modulation, we have chosen $k=0.1$ in Fig. \ref{fig:1anew} to ensure that all pulse shapes are clearly visible within the same plot. Conversely, in Fig. \ref{fig:1bnew}, we have opted for $k=1$, which is the parameter employed in the results for quadratically phased pulses in Fig. \ref{fig:2newn}. 
The values selected for $k=1$ and $\Omega=1$ in this paper are considerably smaller compared to the typical values used in experimental studies \cite{Karpinski2021,torres2011space,Karpi_ski_2016,So_nicki_2023,Yu_2022,Zhu_2022,Bennett_2001,sosnicki2020aperiodic,Karpinski2017}. 
They suggest the potential direction of future experimental efforts to realize the benefits of chirping in single-photon spectroscopy.

Fig. \ref{fig:1cnew} displays the frequency domain pulse shapes of a real exponentially decaying pulse compared to those of a quadratically phase-modulated pulse, illustrating a significant alteration in the frequency domain pulse profile. We delve into its effects on QFI in Sec. \ref{exponential}.

\section{Real Gaussian Pulse} \label{sec:phase}

We first consider a real Gaussian temporal profile
\begin{equation}
    \label{eq:GaussianRealPulse}
    \xi_R(t)=\frac{1}{(2\pi T^2)^{1/4}} e^{-{t^2}/{4 T^2}}, 
\end{equation}
where $T$ is the pulse duration and focus on linear, quadratic, and sinusoidal temporal phases that modulate this real pulse.

\subsection{Linear Temporal Phase}\label{sec:lineaphase1}

A linear phase in the time domain is given by\footnote{Experimentally, a linear temporal phase across a pulse can be achieved using a ``time prism''}~\cite{Karpinski2021}
\begin{equation}
    \phi(t)= \alpha t.
    \end{equation}
This is equivalent to introducing a nonzero detuning and can be identified with the $\Delta$ in the frequency domain matter characteristic function $f(\omega)$ in Eq.~\eqref{eq:char}.

\begin{figure}[ht]
    \centering
    \includegraphics[width=0.8\columnwidth]{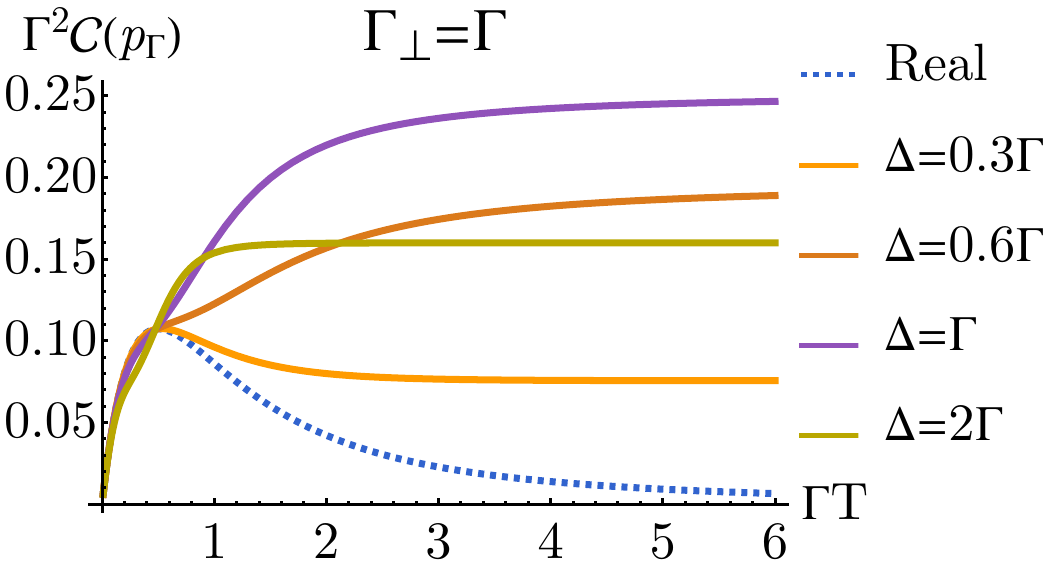}	
    \caption{ Classical contributions to the total QFI for a Gaussian pulse subjected to linear phase modulation, compared to those of a real pulse, $\Gamma_\perp = \Gamma$.} 
    \label{fig:13anew}
\end{figure}

In Fig. \ref{fig:13anew}, we present the classical contribution $\mathcal{C}(p_{\theta})$ to the total QFI $\mathcal{Q}(\rho_{\theta})$, obtained numerically using Eqs. (\ref{eq:pgamma}) and (\ref{eq:dpgamma}) in conjunction with Eq. (\ref{eq:QFI_1ph_normalized}). The plot depicts the classical contribution as a function of pulse durations of the incoming pulse for different detunings $\Delta$. We have chosen $\Gamma_{\perp}=\Gamma$ to highlight a particular value for which the modulated pulses retain more information than the real one.

Exploring various $\Gamma_{\perp}$ values reveals that in the fluorescence detection limit of $\mathrm{max}(\Gamma,\Gamma_{\perp})t\gg 1$, and for T-E coupling strength values within the range $0 < \Gamma_\perp \lesssim 3\Gamma/2$, the classical contribution of the total QFI is higher when a linearly phased pulse is applied to a real Gaussian profile. While this also holds for an exponentially decaying pulse from Eq. (\ref{CFI-Decay}), we do not present the plots here for brevity.
In this limit of T-E coupling strength, the quantum contribution of a linearly phased pulse is always reduced with respect to a real pulse,  for both Gaussian and exponentially decaying profiles. 
\begin{figure*}[th]
    \subfloat[ \label{fig:2anewn}]{
	\includegraphics[width=0.24\textwidth]{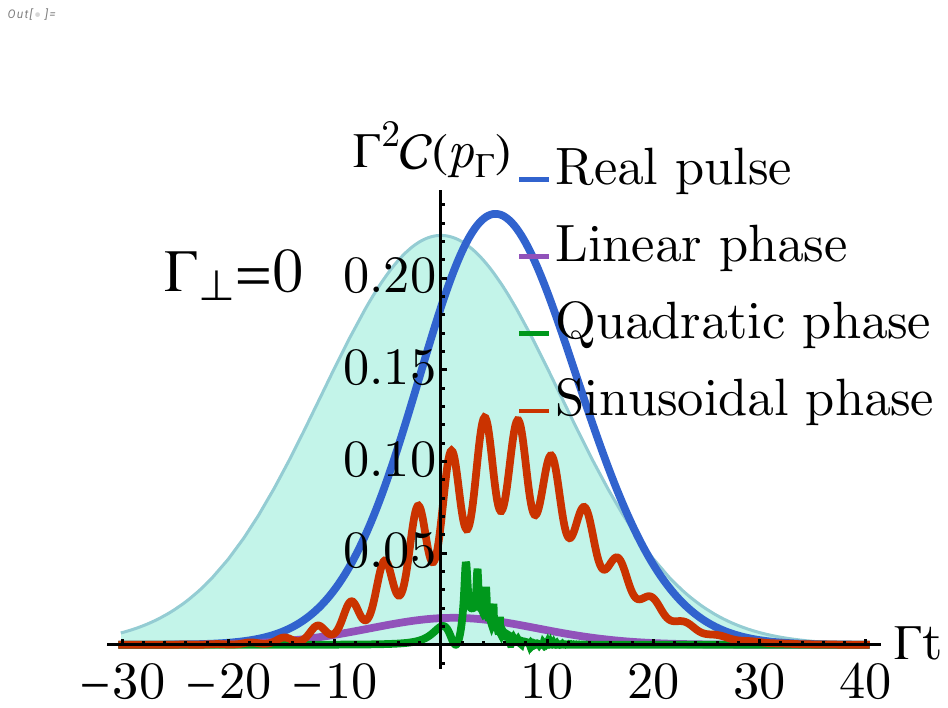}
	}
    \subfloat[ \label{fig:2bnewn}]{
	\includegraphics[width=0.24\textwidth]{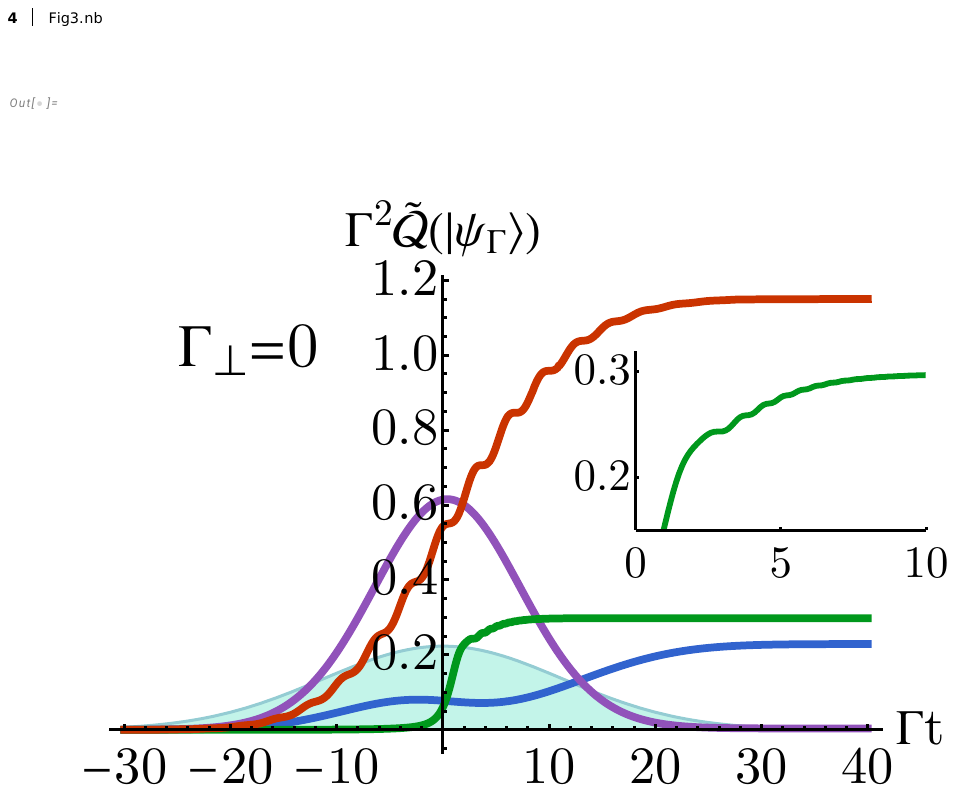}
	}
   \subfloat[\label{fig:2cnewn}]{
	\includegraphics[width=0.24\textwidth]{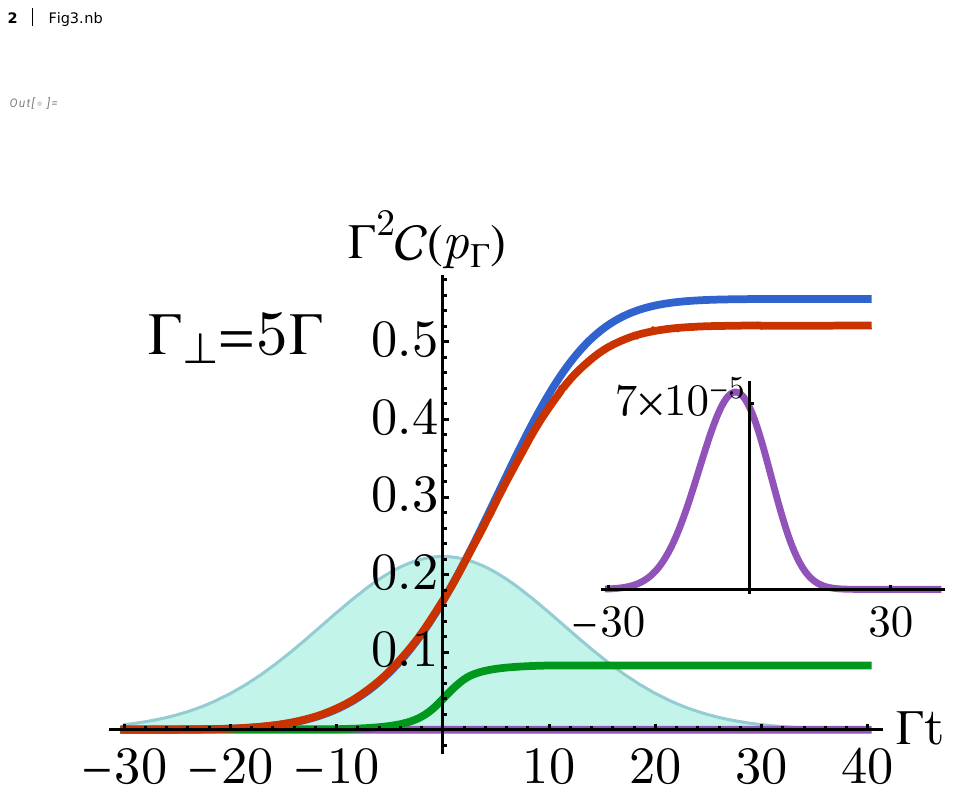}
	}
    \subfloat[ \label{fig:2dnewn}]{
	\includegraphics[width=0.24\textwidth]{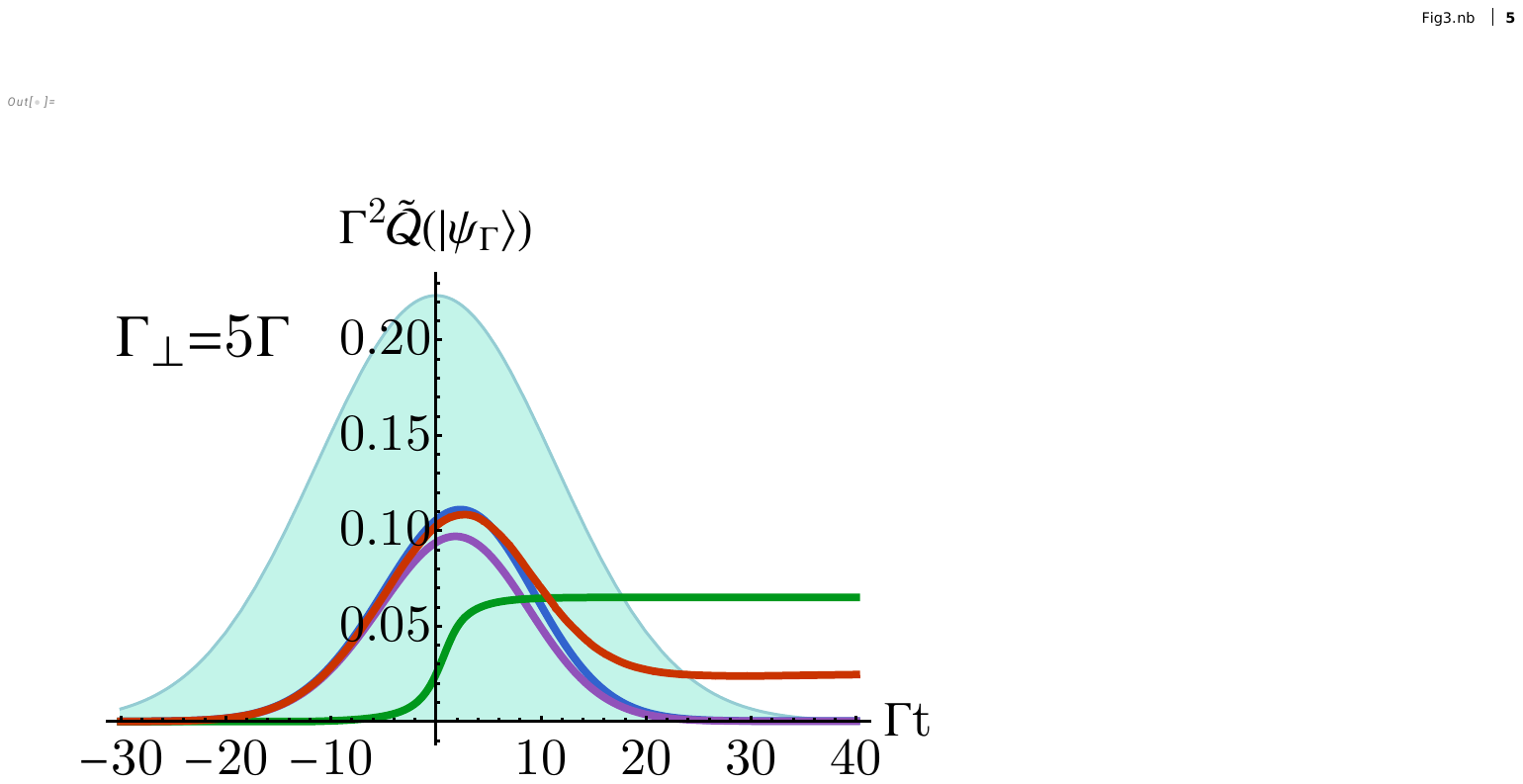}
	}
    \caption{Dimensionless finite-time classical ((a), (c)) and quantum contributions ((b), (d)) to the total QFI for a Gaussian pulse subjected to linear, quadratic, and sinusoidal phase modulations. Parameter values set to $\Delta=\Gamma$, $k=1$, $\Omega=1$, $\Gamma T = 8$. $\Gamma_\perp=0$ in (a),(b) and $\Gamma_\perp=5 \Gamma$ in (c),(d). The shaded purple region represents the modulus squared of the temporal amplitude as a visual aid; it is not drawn to scale on the vertical axis.}
    \label{fig:2newn}
\end{figure*}

For larger values of the T-E coupling strength (${{\Gamma }_{\bot }}\gtrsim 3\Gamma/2 $), the effect of linear phase is evident as a reduction in the quantum and classical contributions of the total QFI compared to those related to a real pulse (see Figs. \ref{fig:CFI5} and \ref{QFIS*(1-p)}).

In Fig. \ref{fig:2newn}, we display the classical and quantum contributions of the total QFI as a function of the detection time $t$ for the Gaussian pulse when it is subjected to linear phase modulation (purple lines). Except for an increase in the quantum contribution of the linearly phased pulse for $\Gamma_\perp=0$ (in Fig. \ref{fig:2bnewn}), this phase always decreases the obtained information for both $\Gamma_\perp=5 \Gamma$ and $\Gamma_\perp=0$, compared to the real pulse. 

In the scenario of perfect T-P coupling, employing a perfectly resonant wave-packet yields a greater amount of information asymptotically, and adding a linear phase consistently decreases the QFI. However, in the finite-time regime, this does not hold (see Fig. \ref{fig:2bnewn}). 
This is evident from Fig. \ref{fig:2bnewn}, where we observe that the quantum contribution $\Gamma^2 \tilde{\mathcal{Q}}(\ket{\psi_{\Gamma}})$ for the linearly phase-modulated pulse grows significantly faster as a function of the detection time $t$ than that of the real pulse. This implies that for small and intermediate detection time regimes, the linearly phase-modulated pulse may be used as a more sensitive probe for $\Gamma$-estimation.
It is evident that $\Gamma^2 \widetilde{\mathcal{Q}}(\left| {{\psi }{\theta }} \right\rangle )+\Gamma^2 \mathcal{C}({{p}{\theta }})$ (the sum of dimensionless values in Figs. \ref{fig:2anewn} and \ref{fig:2bnewn}) is greater for the linear phase pulse compared to the real pulse.

\subsection{Quadratic and Sinusoidal Temporal Phases} \label{sec:quad}

We next consider a quadratic phase in the time domain
\begin{equation} 
    \label{eq:quadphase}
\phi (t)= k {\Gamma}^2 t^2,
\end{equation} 
where $k$ is a real, dimensionless chirping strength. 
For a real Gaussian pulse of Eq.~\eqref{eq:GaussianRealPulse}, with frequency bandwidth $\sigma_\omega=1/{(2\Gamma T)}$, the corresponding phase modulated pulse in the frequency domain is still Gaussian and has a bandwidth given by $\sigma'_{\omega }={\sqrt{1+16{{k}^{2}}\Gamma^4T^{4}}}\sigma_\omega$. 
The role of quadratic temporal phase modulation in the estimation of time and frequency shifts in generalized Hong-Ou-Mandel interferometry was studied in Ref.~\cite{Fabre2021}.

A sinusoidal phase in the time domain
\begin{equation}
\phi(t) = \sin (\Omega \Gamma t), 
\end{equation}
where $\Omega$ is a dimensionless modulation strength, not only modifies the bandwidth $\sigma _{\omega }$ but also introduces sidebands, altering the pulse shape in the frequency domain. 


\subsubsection{Finite time}

The finite-time classical and quantum contributions to the total QFI are depicted in Fig \ref{fig:2newn}. 
The real temporal pulse provides more classical contribution for both coupling strengths of $\Gamma_\perp = 0$ and $\Gamma_\perp = 5\Gamma$. 
Conversely, a quadratically phased pulse provides more quantum contribution than a real pulse.

For the quadratically phased pulse in the absence of T-E coupling ($\Gamma_{\perp}=0$), oscillations are observed in the excitation probability (not shown here), as well as in the classical and quantum contributions of the total QFI (Figs. \ref{fig:2anewn} and \ref{fig:2bnewn} respectively). These oscillations are absent in the presence of relatively strong coupling to the environment, i.e., $\Gamma_\perp = 5\Gamma$ (Figs. \ref{fig:2cnewn} and \ref{fig:2dnewn}). 
It is important to note that all these behaviors depend on the value of $\Gamma T$. For instance, in the case of $\Gamma T = 2$ and $\Gamma_\perp = 0$ (not shown here), the quadratically phased pulse preserves more information (in the quantum contribution of $\Gamma^2 \mathcal{Q}\left( \ket{\psi_\Gamma} \right)$) up to a specific time, after which the situation is reversed. 
Therefore, in the finite-time regime, there is no general rule as to whether a quadratically phased Gaussian pulse is beneficial for $\Gamma$-estimation. 
One needs to examine each case using the expressions provided in Appendix \ref{App:A} for each contribution of the finite-time QFI. 
This holds for both sinusoidal and linear phases as well.
 
For the quadratic temporal phase, the real temporal pulse provides a higher classical contribution and a lower quantum contribution compared to a sinusoidally modulated pulse. See Figs \ref{fig:2newn}.
However, both these contributions (classical and quantum contributions) exceed those compared to the quadratic one. In both zero and large coupling to the environment (i.e., both $\Gamma_\perp = 0$ and $\Gamma_\perp = 5 \Gamma$), the classical contribution of the sinusoidal phase is lower than that of the real pulse (Figs. \ref{fig:2anewn}, \ref{fig:2cnewn}). The quantum contribution is higher than that of a real pulse in the case of $\Gamma_\perp = 0$ (Fig. \ref{fig:2bnewn}), and in $\Gamma_\perp = 5 \Gamma$ (Fig. \ref{fig:2dnewn}), it surpasses that of the real pulse at certain times, although the quadratic phase is doing better than both real and sinusoidal phase at this time.

It is worth noting that a linearly phased pulse retains the highest amount of information in the quantum contribution (see Fig. \ref{fig:2bnewn}) at the beginning of the TLS-pulse interaction for $\Gamma_\perp = 0$.
We emphasize once again that these results are presented for $\Gamma T=8$, and different pulse durations may exhibit different behaviors.

\subsubsection{\texorpdfstring{$t \gg 1/\mathrm{max}(\Gamma,\Gamma_{\perp})$}{}}

Analytical expressions for the classical and quantum contributions to the FI of a quadratically phase-modulated Gaussian pulse, 
as described by Eqns.~\eqref{eq:GaussianRealPulse} and \eqref{eq:quadphase}, at asymptotically long times $t \gg 1/\mathrm{max}(\Gamma,\Gamma_{\perp})$ is
presented in Appendix \ref{R vs PM}.
As Eqns. (\ref{eq:CFI_gau_A}) and (\ref{eq:QFI_gau_A}) show, the sole pulse characteristic they depend on is its bandwidth.
As in the classical case, quadratic phase modulation thus provides a valuable technique for manipulating the bandwidth of a pulse for quantum light spectroscopy.

\begin{figure*}[ht]
    \subfloat[\label{fig:CFI5}]{
        \includegraphics[width=0.32\textwidth]{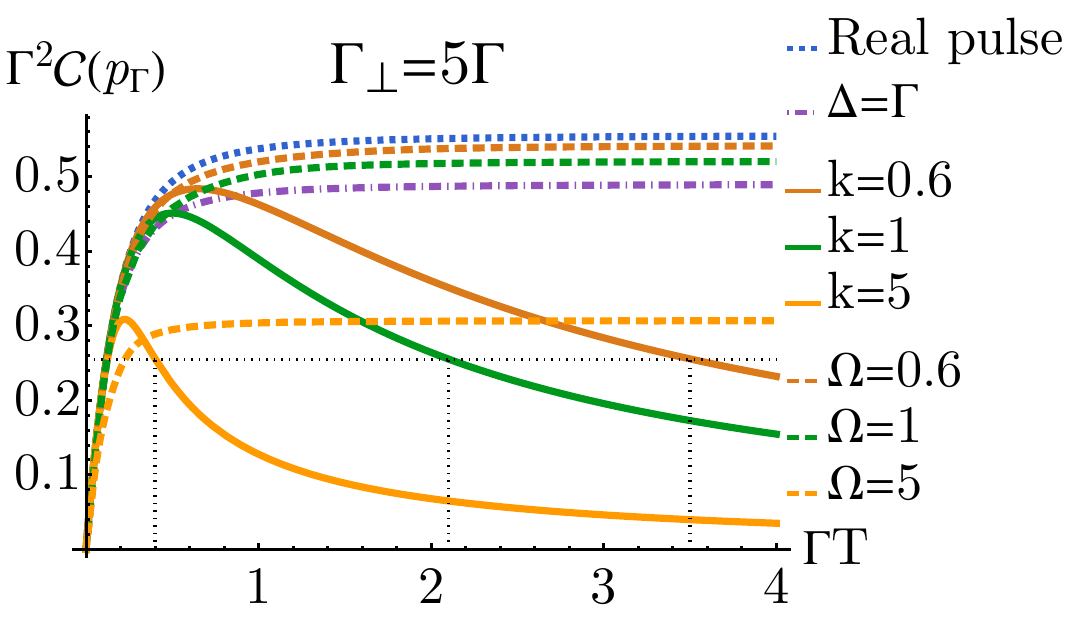}
        }
	\subfloat[\label{QFIS*(1-p)}]{
		\includegraphics[width=0.32\textwidth]{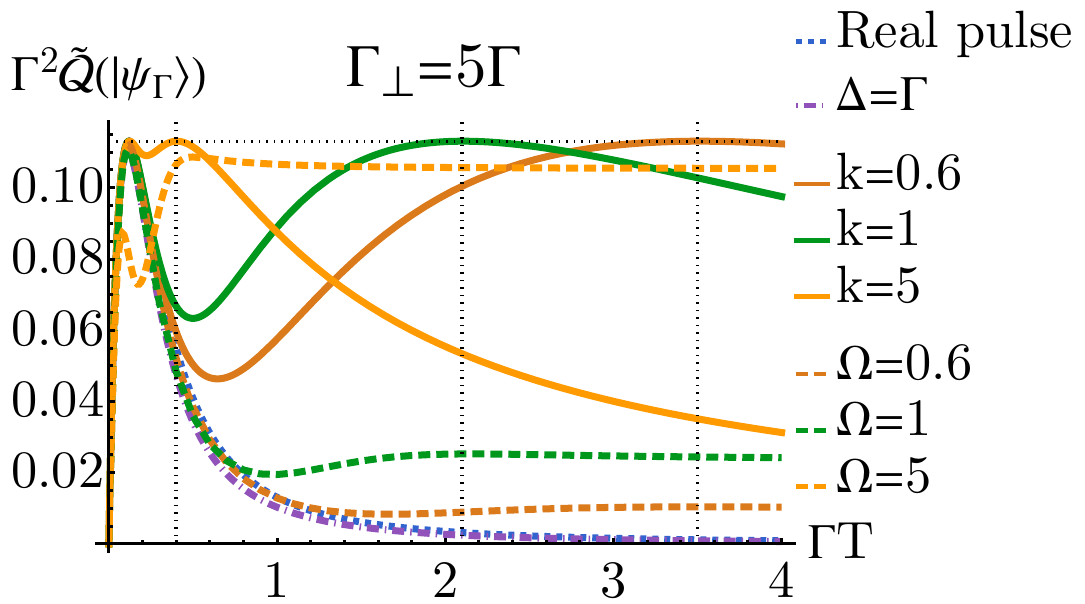}
		}
     \subfloat[\label{QFI_sin_g0}]{
		\includegraphics[width=0.32\textwidth]{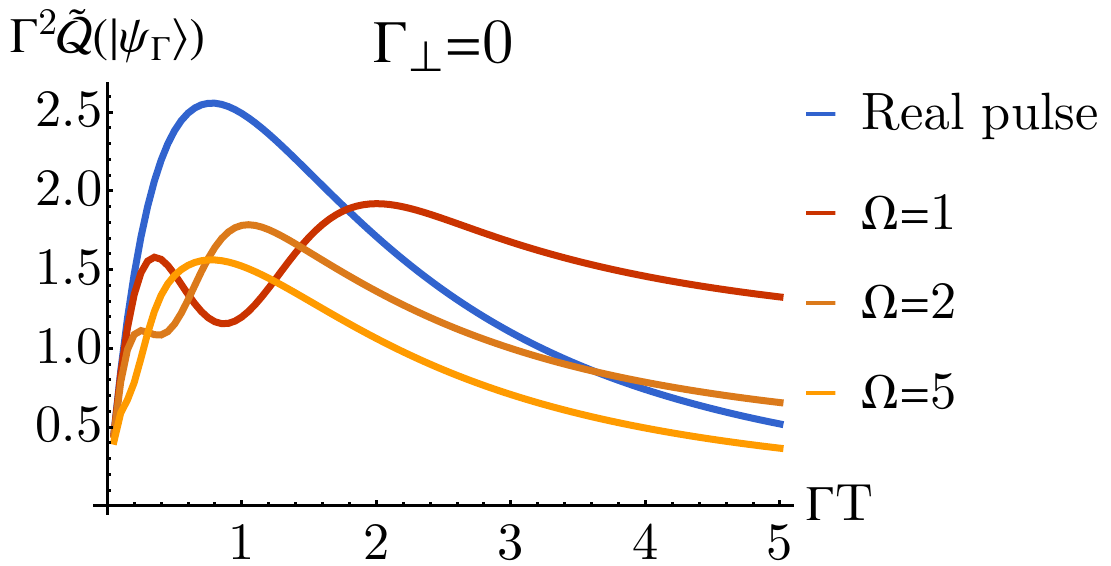}
		}    
    \caption{Asymptotic (a) classical and (b) quantum contributions to the total QFI for a Gaussian pulse subjected to linear (dash-dotted line), quadratic (solid lines), 
    			and sinusoidal (dashed lines) phase modulations separately for $\Gamma_\perp = 5\Gamma$. 
			(c) The total QFI for a Gaussian pulse subjected to sinusoidal phase modulation for $\Gamma_\perp = 0$. 
			The real pulse corresponds to $\Delta=0$, $k=0$ and $\Omega=0$. }
	\label{fig:QFI}
\end{figure*}

Given the cumbersome expressions in Appendix~ \ref{R vs PM}, we plot the classical contributions in Fig. \ref{fig:CFI5} for the illustrative value of $\Gamma_\perp=5\Gamma$.
As $k$ decreases, these contributions increase. Unsurprisingly, the contribution for $k=0$ must match that of the unmodulated real pulse. 
As Eq.~\eqref{eq:CFI_gau_A} also shows, the classical (and quantum) contributions depend on the ratio $\gamma = \Gamma_{\perp}/\Gamma$. 
Indeed, for $\gamma=1$, the chirped pulse provides more classical information than the unchirped, real pulse.

The quantum contribution, depicted in Fig. \ref{QFIS*(1-p)} for different $k$ values and  $\Gamma_\perp=5\Gamma$, exhibits richer behaviour.
As the bottom inset of Fig. \ref{fig:1bnew} displays, for $k>0$ the same value of $\sigma_{\omega }$ corresponds to two different $\Gamma T$.
This leads to the two peaks in the solid curves in Fig.~\eqref{QFIS*(1-p)} indicated by the dotted line to specify the $\Gamma T$ values. This also applies to the classical contribution, which manifests as equal values twice along each solid curve in Fig. \ref{fig:CFI5}, as indicated by the dotted line.

As it is evident, the relative performance of a quadratically phase-modulated -- 
in terms of its classical and quantum contributions to the total QFI depends on $\gamma$ and $\Gamma T$. This also applies to a sinusoidally phase-modulated, as we discuss next.

The effects of sinusoidal phase modulation on the real Gaussian pulse are illustrated in Figs. \ref{fig:CFI5} and \ref{QFIS*(1-p)} in the presence of environmental loss with $\Gamma_\perp=5\Gamma$. As it is shown in Fig. \ref{fig:1anew}, the pulse shape ${{\left| \tilde{\xi} (\omega ) \right|}}$, and not only its width is altered in comparison to the real pulse when sinusoidal phase modulation is applied. Consequently, we do not expect to obtain identical results using the real pulse with the same frequency domain bandwidth $\sigma_{\omega }$. 

For $\Gamma_\perp=5 \Gamma$, an increase in $\Omega$ reduces the classical contributions of the phased pulse. 
However, the quantum contribution, which arises from distortions in the pulse wave-packet, exhibits various behaviors with changes in $\Omega$. 
The behaviour of both classical and quantum contributions cannot be solely explained by alterations in the frequency domain bandwidth. 

From Fig. \ref{fig:1bnew} (the red curve), it is apparent that the bandwidth remains constant at $\Gamma T \gtrsim 2$ for $\Omega=1$, yet $\Gamma^2\tilde{\mathcal{Q}}(\ket{\psi_\Gamma})$ of $\Omega=1$ (Fig. \ref{QFIS*(1-p)} dashed green curve) does not contain the same amount of information for $\Gamma T \gtrsim 1$. 
The dashed lines in \ref{fig:CFI5} and \ref{QFIS*(1-p)} seem pretty flat after $\Gamma T \gtrsim 2$.

\begin{figure*}[ht]
    \subfloat[\label{fig:3anew}]{
	\includegraphics[width=0.28\textwidth]{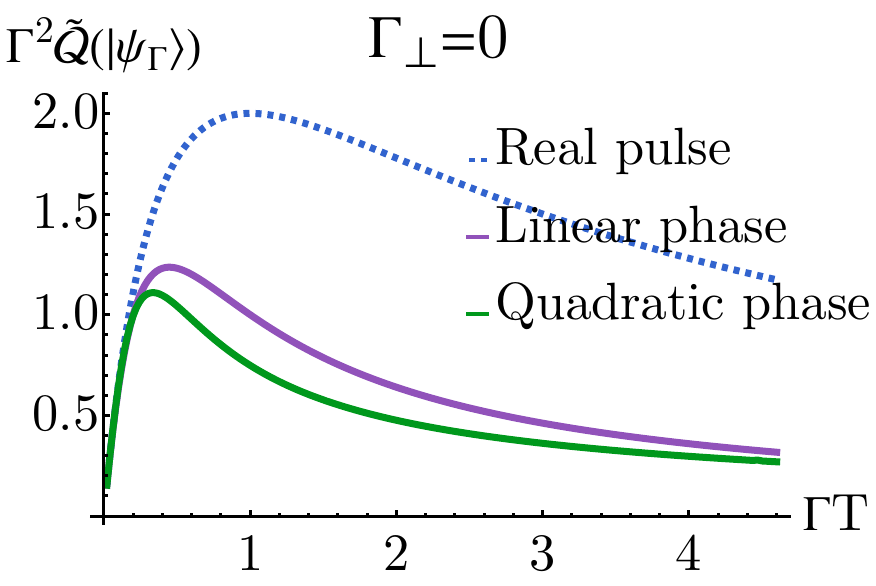}
	}
    \subfloat[\label{fig:3bnew}]{
	\includegraphics[width=0.28\textwidth]{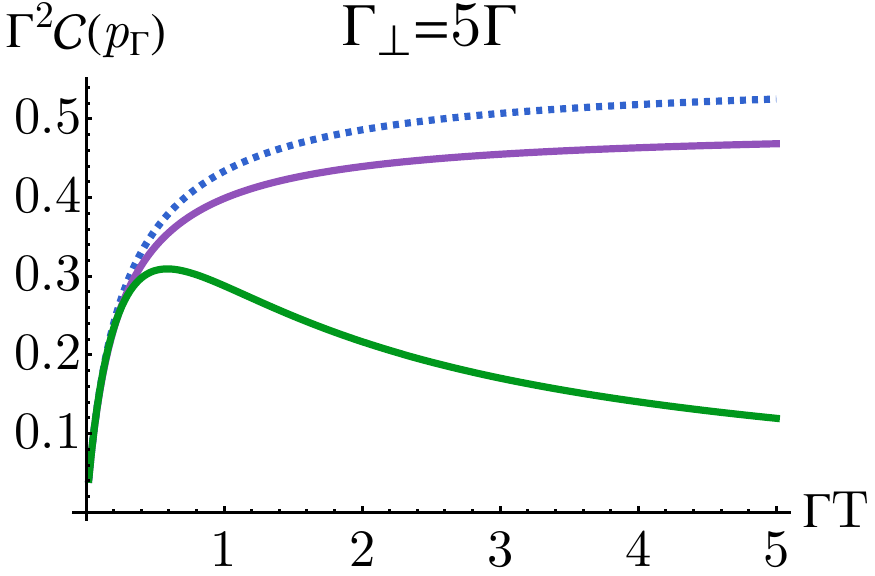}
	}
    \subfloat[\label{fig:3cnew}]{
	\includegraphics[width=0.28\textwidth]{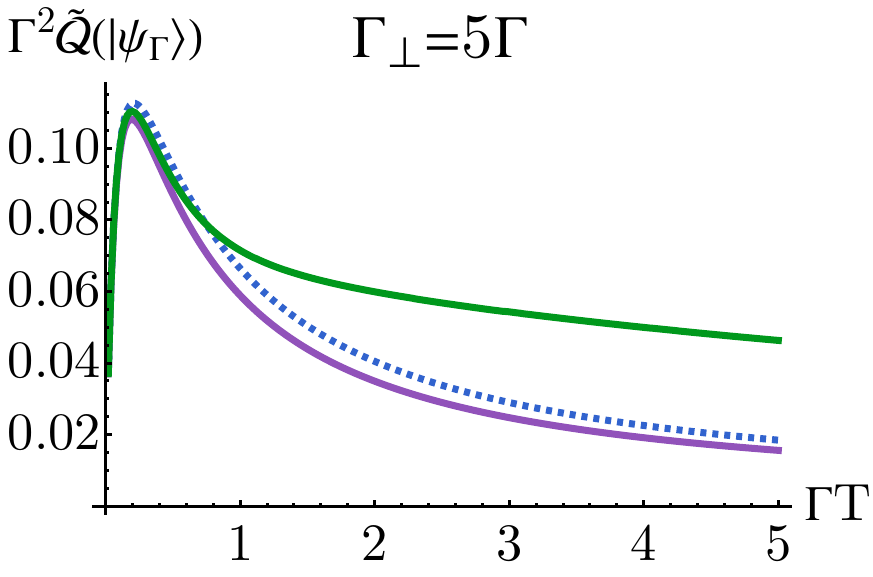}
	}
    \caption{  (a),(c)  Asymptotic quantum contribution and (b) asymptotic classical contribution to the total QFI for the exponentially decaying pulse, compared to those associated with the linearly and quadratically phase-modulated pulses. The parameter settings include $\Gamma_\perp=0$ in (a) ,$\Gamma_\perp=5 \Gamma$ in (b),(c). We set $k=1$ for the quadratic phase and $\Delta=\Gamma$ for the linear phase.} 
    \label{fig:3new}
\end{figure*}

To clarify this point, we illustrate the asymptotic QFI in the case of $\Gamma_\perp=0$ in Fig. \ref{QFI_sin_g0}. It is important to note that in the asymptotic time when $\Gamma_\perp=0$, the classical contribution is zero, as both terms of $p_\theta$ in Eq. (\ref{16}) become zero. One can observe from Fig. \ref{QFI_sin_g0} that, for $\Omega=1$ at $\Gamma T \gtrsim 1$, the QFI exhibits various values, even though the frequency bandwidth (shown in Fig. \ref{fig:1bnew}) remains constant. Fig. \ref{QFI_sin_g0} illustrates the changes in the quantum contribution resulting from alterations in the $\Omega$ values.

\section{Exponentially Decaying Pulse} \label{exponential}

In this section, we examine an exponentially decaying pulse as the real profile of the pulse, given by
\begin{align}
    \xi_{R} (t)={1}/{\sqrt{T}} e^{-{t}/{2T} } \Theta(t),
\end{align}
and apply both linear and quadratic phase modulations to it.
In contrast to the real Gaussian pulse, in the case of an exponentially decaying real pulse, applying quadratic phase modulation does not just affect the frequency bandwidth; it drastically changes the frequency shape of the pulse. As shown in Fig. \ref{fig:1cnew}, the pulse shape in the frequency domain lacks symmetry, necessitating consideration of the second term of Eq. (\ref{eq:qfipure}) to calculate the quantum contribution of the QFI. Adding both linear and quadratic phases results in various changes in both classical and quantum contributions, which will be discussed in the following subsections.

\subsection{Linear  Temporal Phase}

In Fig. \ref{fig:3new}, we display the asymptotic classical and quantum contributions of the total QFI (Appendix \ref{App:decaying}, Eqs. \ref{CFI-Decay} and \ref{QFI-Decay} respectively) using the linearly phased exponentially decaying pulse for both zero and relatively strong coupling to the environment ($\Gamma_\perp=5\Gamma$). This figure also shows the asymptotic classical and quantum contributions of both quadratically phased and real pulses.
For vanishing T-E coupling i.e. $\Gamma_{\perp}=0$, a linearly phased pulse retains less information compared to a real-valued pulse, although it retains more information than a quadratically phased pulse (see Fig. \ref{fig:3anew}). When there is non-zero coupling to the environment within the range $0 < \Gamma_\perp \lesssim 3\Gamma/2$, the classical contribution of the linearly phased pulse surpasses that of a real pulse (not depicted here). However, if ${\Gamma }_{\bot }=5\Gamma $, the classical contribution (depicted in Fig. \ref{fig:3bnew}) shows that a linear phase retains less information compared to a real pulse but more than a quadratically phased pulse. In terms of the quantum contribution (refer to Fig. \ref{fig:3cnew}), a linearly phased pulse retains less information than both real and quadratically phased pulses. The described behaviors of classical and quantum contributions remain consistent within the ranges $ \Gamma_\perp \gtrsim 3\Gamma/2$.

The finite-time classical and quantum contributions of the total QFI are depicted in Fig. \ref{fig:4new} at a fixed pulse duration. At this specific pulse duration ($\Gamma T=4$), when ${\Gamma }{\bot }=0$, a linearly phased pulse exhibits oscillatory behavior in its classical contribution at the beginning of the TLS-light interaction. Apart from these oscillations, the amount of information obtained from a linearly phased pulse is consistently less than that from a real pulse but more than that from a quadratically phased pulse. This observation holds for the classical contribution at ${\Gamma }{\bot }=5\Gamma$ as well. 
For the corresponding quantum contribution, after a certain duration of the interaction, the information obtained from the linearly phased pulse is less than that from both the real and the quadratically phased pulses.

Note that these behaviors are also dependent on the duration of the initial pulse, captured by the parameter by $\Gamma T$. For instance, an increase in the quantum contribution of the linearly phased pulse compared to the real pulse is observable in the exponentially decaying pulse when $\Gamma T=8$ and $\Gamma_\perp=0$, whereas this increase is not evident at $\Gamma T=4$.

\subsection{Quadratic Temporal Phase }

The change in the frequency domain pulse shape resulting from quadratic time phase modulation applied to an exponentially decaying pulse is illustrated in Fig. \ref{fig:1cnew}.
Quadratic phase modulation does not just alter the frequency domain bandwidth, as observed with a Gaussian real pulse. Instead, it drastically modifies the frequency domain pulse shape.
We first explore the effect of this change in the asymptotic time QFI represented in Fig. \ref{fig:3new}. 
For $\Gamma_\perp=0$, as depicted in Fig. \ref{fig:3anew}, quadratic phase modulation consistently decreases the obtained information. Interestingly, this decrease is observed in the classical contribution for $\Gamma_\perp=5 \Gamma$ as well (seen in Fig. \ref{fig:3bnew}). Conversely, the quantum contribution implies that the quadratically phase-modulated wave-packet retains more information when $\Gamma T \geq 0.8$ for strong coupling of the TLS to the environment, represented by $\Gamma_\perp=5\Gamma$ (visible in Fig. \ref{fig:3cnew}). 

\begin{figure*}[ht]
    \subfloat[\label{fig:4bnew}]{
	\includegraphics[width=0.24\textwidth]{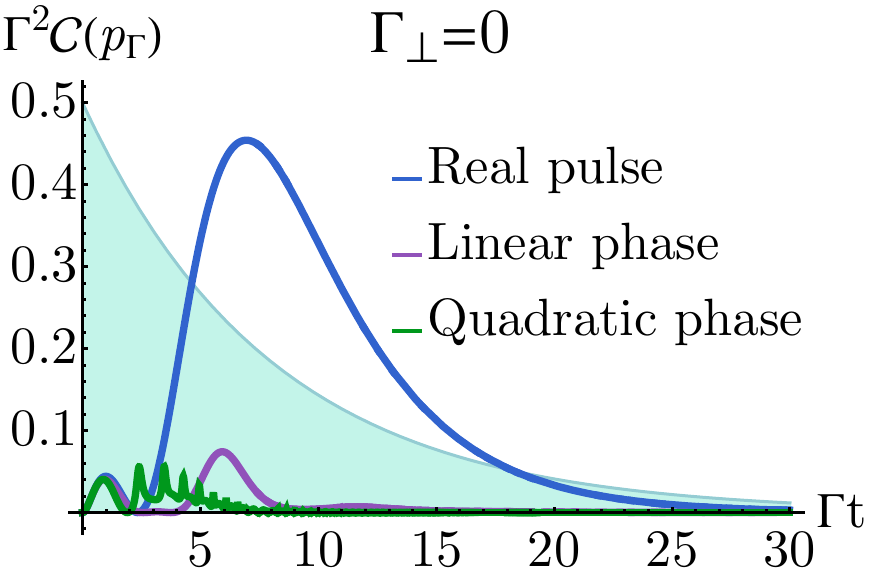}
	}
    \subfloat[\label{fig:4cnew}]{
	\includegraphics[width=0.24\textwidth]{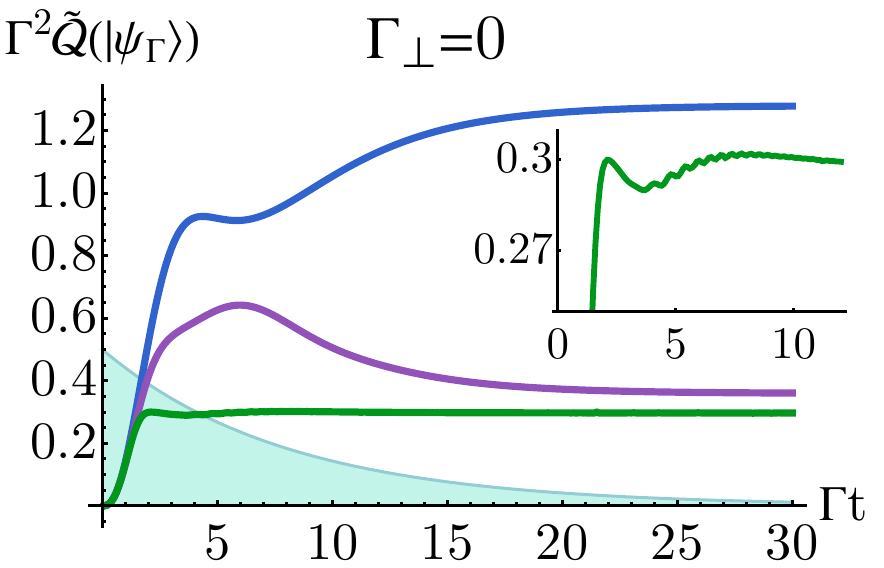}
	}
    \subfloat[\label{fig:4dnew}]{
	  \includegraphics[width=0.24\textwidth]{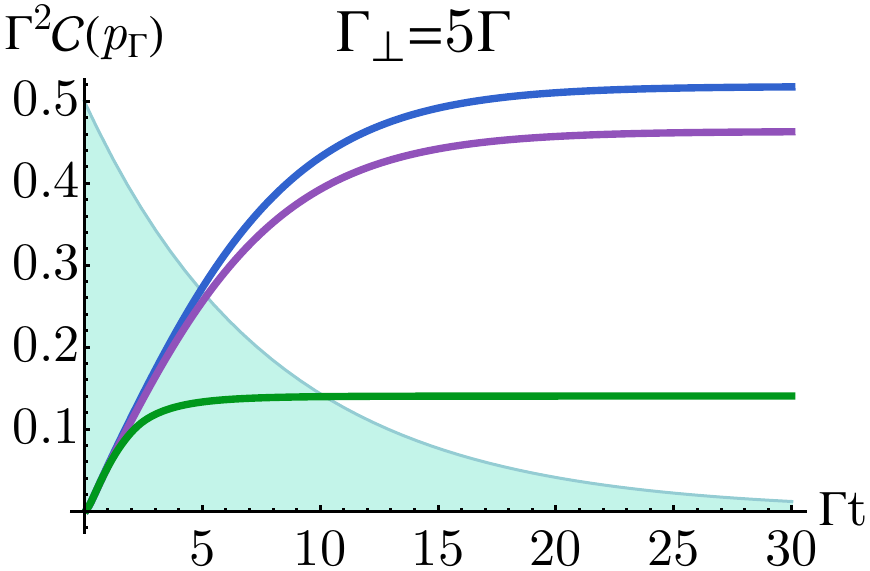}
	  }
    \subfloat[\label{fig:4enew}]{
	  \includegraphics[width=0.24\textwidth]{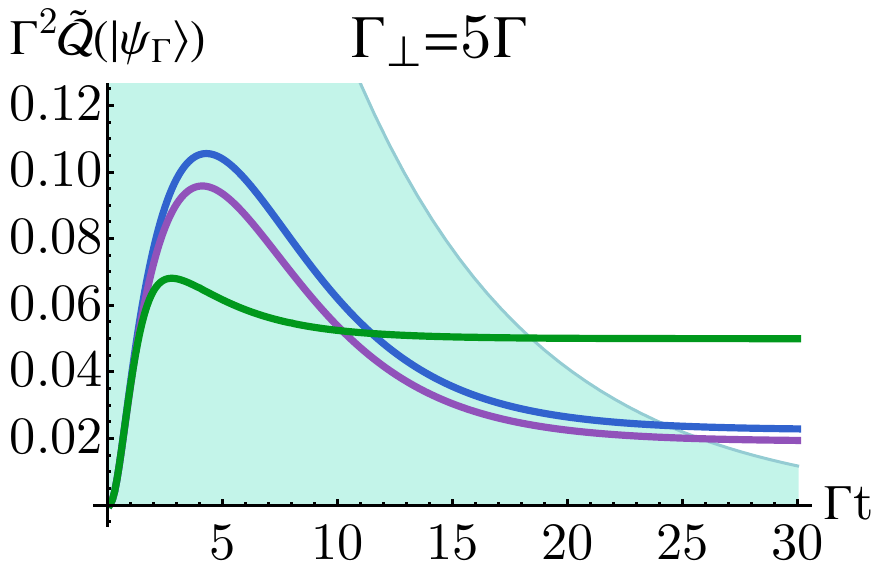}
	  }
    \caption{ (a),(c) Finite-time classical contribution and (b),(d) finite time quantum contribution to the total QFI for the exponentially decaying pulse, compared to those associated with the linearly and quadratically phase-modulated pulses. The parameter settings include $\Gamma T=4$,  $\Gamma_\perp=0$ in (a),(b) ,$\Gamma_\perp=5 \Gamma$ in (c),(d). We set $k=1$ for the quadratic phase and $\Delta=\Gamma$ for the linear phase.} 
    \label{fig:4new}
\end{figure*}

The finite time classical and quantum contributions of the total QFI are depicted in Fig. \ref{fig:4new} for two distinct coupling strengths to the environment: $\Gamma_\perp=0$ and $\Gamma_\perp=5\Gamma$.
The oscillations observed for $\Gamma_\perp=0$ in the probability of the TLS excitation (not shown here), the classical and quantum contributions using the quadratically phased pulse in Figs. \ref{fig:4bnew}, \ref{fig:4cnew} are absent in the case of strong coupling to the environment indicated by $\Gamma_\perp=5 \Gamma$ (Fig. \ref{fig:4dnew}, \ref{fig:4enew}).
In both values of T-E coupling, the quadratically phased pulse retains the lowest information in classical contribution compared to the linearly phased and real pulses. This observation holds for the quantum contribution of $\Gamma_\perp=0$ as well (seen in Fig. \ref{fig:4cnew}).
Significantly, the benefits of quadratic phase modulation become apparent in the quantum contribution, in the scenario with strong coupling to the environment, $\Gamma_\perp=5\Gamma$. This outcome is depicted in Fig. \ref{fig:4enew} in comparison to the cases of linearly phased and real pulses. Similar to the Gaussian pulse, after a certain duration of the TLS-pulse interaction, the quadratically phased pulse retains more information than both the real and linearly phased pulses.

\section{Optimal and Near-Optimal Measurements} \label{sec:Measurement}

Having obtained the fundamental limits of estimating the inverse lifetime  $\Gamma$ of a single TLS, we now seek the limits that can be attained using feasible detection strategies.
For single-parameter estimation, at least one optimal POVM whose CFI attains the QFI exists, but it may depend on the true value of the parameter.
Thus, the QCRB is practically achieved by measuring multiple copies of the state and employing a two-step adaptive procedure, i.e. devoting a sublinear portion of the experiment repetitions to compute a rough estimator and then applying the optimal measurement corresponding to the estimated value~\cite{Barndorff-Nielsen2000}.

Optimal detection for the outgoing P state in Eq.~(\ref{eq:outgoing_state}) comprises two distinct parts~\cite{albarelli2022fundamental}: photon loss measurement, effected by POVM element $\Pi_0 = \ket{0^{\mathrm{P}}}\bra{0^{\mathrm{P}}}$ that corresponds to the vacuum component, as well as POVM elements that act on the single-photon component $\Pi_{1,s} = \int d\tau d\tau^\prime\,\Pi_{1,s}(\tau,\tau^\prime)a^{\dag}(\tau)\ket{0^{\mathrm{P}}}\bra{0^{\mathrm{P}}}a(\tau^\prime)$.
The corresponding probabilities can be evaluated as $p_0 = \mathrm{Tr}[\rho_{\theta}\Pi_0] = p_{\theta}$, and $p_{1,s} = \mathrm{Tr}[\rho_{\theta}\Pi_{1,s}] = (1-p_{\theta})\langle\psi_{\theta}|\Pi_{1,s}|\psi_{\theta}\rangle$.
Using the chain rule for Fisher information, the associated CFI is then 
\begin{equation}
    \mathcal{C}(\theta|\{p_0,p_{1,s}\}) = \mathcal{C}(\theta|\{p_{\theta},1-p_{\theta}\}) + (1-p_{\theta})\mathcal{C}(\theta|\{p_{1|s}\}), 
\end{equation}
where $p_{1|s} = \langle\psi_{\theta}|\Pi_{1,s}|\psi_{\theta}\rangle$ is the conditional probability of outcome $s$ in the one-photon component $\ket{\psi_{\theta}}$.
Therefore, the classical contribution to the QFI in Eq.~\eqref{eq:QFI_1ph_normalized} can always be saturated using the POVM element $\Pi_0$ that perfectly distinguishes the vacuum component of $\rho_{\theta}$ from the single-photon component.

The quantum contribution must coincide with the CFI associated with the optimal basis corresponding to the conditional one-photon state $\ket{\psi_{\theta}}$.
In the following, we will discuss optimal estimation techniques corresponding to the conditional state $\ket{\psi_{\theta}}$, with the understanding that QFI corresponding to the full state $\rho_{\theta}$ can be attained by supplementing the optimal POVM elements for $\ket{\psi_{\theta}}$ with the loss operator $\Pi_0$. 

A particular optimal measurement for pure state models is any POVM that includes a projector onto the pure state itself~\cite{albarelli2022fundamental}; however, it was recently pointed out that there are some subtleties in the implementation of the two-step adaptive procedure~\cite{Girotti2023}.
More generally, the choice of a QCRB-saturating measurement for pure state models is not unique, and there is an infinite number of optimal POVMs.
Between these, a certain $2$-outcome POVM was shown to be least susceptible to small, but arbitrary, measurement noise, as quantified by the recently introduced Fisher information measurement noise susceptibility~\cite{kurdzialek2022measurement}.

In the following, we will focus on more realistic strategies, based on projection on orthonormal temporal modes, as a more realistic measurement strategy. We will also specialize our discussion to the specific choice of $\theta=\Gamma$ for clarity of description, although the formulation is equally applicable to other Hamiltonian parameters.

\subsection{\texorpdfstring{$\langle \psi_{\Gamma} | \partial_{\Gamma} \psi_{\Gamma} \rangle = 0$}{}}\label{sub:overlapzero}

The probabilities of detecting the outgoing single photon state  $\ket{\psi_{\Gamma}}$ in the discrete basis modes $\{\ket{a_j} = a_j^{\dag}\ket{0^{\mathrm{P}}}\}$~
(for a fuller description, see Appendix \ref{app:modaldecom}), and their $\Gamma$-derivatives, are 
\begin{align}
     p_j = |\langle a_j | \psi_{\Gamma} \rangle |^2 = \begin{cases}
      \frac{1}{1-p_{\Gamma}}(1-C_0)^2, &\ j=0 \\
      \frac{1}{1-p_{\Gamma}}(C_j)^2, &  j\in\mathcal{N}.
    \end{cases}
\end{align}
\begin{align}
    \frac{\partial p_j}{\partial \Gamma} =
     \begin{cases}
      -2(1-C_0)D_0^{\Gamma}/(1-p_{\Gamma}) + \frac{p_{\Gamma}}{1-p_{\Gamma}}p_0, &  j=0 \\ \\
      2C_j^g D_j^{\Gamma}/(1-p_{\Gamma}) + \frac{p_{\Gamma}}{1-p_{\Gamma}}p_j, &  j\in\mathcal{N}.
    \end{cases} 
\end{align}
Then the $\Gamma$-CFI associated with the mode-resolved photon counting measurement is
\small
\begin{align}\label{eq:predictedconverge}
    &\mathcal{C}(\{p_j\})\big\vert_{\langle \psi_{\Gamma} | \partial_{\Gamma} \psi_{\Gamma} \rangle = 0} = \sum_j~\frac{1}{p_j}\left( \frac{\partial p_j}{\partial \Gamma}\right)^2 \bigg\vert_{\langle \psi_{\Gamma} | \partial_{\Gamma} \psi_{\Gamma} \rangle = 0} \nonumber\\
    &= \frac{4}{1-p_{\Gamma}}~\sum_j~(D_j^{\Gamma})^2 - \frac{(\partial_{\Gamma}p_{\Gamma})^2}{(1-p_{\Gamma})^2} =   \mathcal{Q}(\rho_{\theta})\bigg\vert_{\langle \psi_{\Gamma} | \partial_{\Gamma} \psi_{\Gamma} \rangle = 0},
\end{align}
\normalsize
where we have utilised the fact that $\mathrm{Im}\left[(1-C_0^{*})D_0 - \sum_j\,C_j^* D_j\right] = 0$ for $\langle \psi_{\Gamma} | \partial_{\Gamma} \psi_{\Gamma} \rangle = 0$. 

What we have demonstrated here is then quite general: $\Gamma$-estimation for an outgoing pulses that satisfy the condition $\langle \psi_{\Gamma} | \partial_{\Gamma} \psi_{\Gamma} \rangle = 0$ is optimal in the fixed complete basis $M = \{ \ket{a_0}\bra{a_0},\ket{a_1}\bra{a_1},\dots \}.$ In Appendix \ref{app:realvec_pulses}, we show that all incoming traveling pulses with symmetrical spectral distribution $|\tilde{\xi}(\omega)|^2$ respect this condition, meaning all modulated pulses in Fig. \ref{fig:1anew} are optimally measured in the fixed basis $M$, in a single, non-adaptive step. In contrast, quadratically modulated exponentially decaying pulses~(see Fig. \ref{fig:1cnew}) violate this condition owing to the asymmetry of their spectral shape, and the outgoing pulses are suboptimal in the fixed basis $M$. 

This result is reminiscent of the use of spatial mode demultiplexing~(SPADE) measurements~\cite{tsang2016quantum,nair2016far,rehacek2017optimal,bisketzi2019quantum} that can be used to exceed the Rayleigh limit for spatially separated incoherent sources, and are directly comparable to the temporal analogue of the same  problem~\cite{mitchell2022quantum} where the optimality of the pulse-envelope basis for exponentially decaying pulses has already been demonstrated for discrimination 
of the spontaneous emission lifetimes of two TLSs.

Mode-resolved photon counting can be achieved using quantum pulse gating~(QPG) techniques~\cite{eckstein2011quantum,donohue2018quantum,de2021effects,ansari2021achieving,garikapati2022programmable} for ultrafast pulses.
With the right toolbox of gating pulses, the optimal measurement for $\Gamma$-estimation is accessible, in principle. 
In practice,  the number of mode projectors that can be implemented in an actual experiment is limited.
The QPG measurement along the mode basis set by the incoming pulse envelope is more accurately represented using the POVM $M_J = \{\ket{a_0}\bra{a_0},\dots,\ket{a_J}\bra{a_J}, \mathds{1}^{\mathrm{P}}-\ket{a_0}\bra{a_0}-\dots\ket{a_J}\bra{a_J}\}$, where $J$ is a number set by practical considerations. 

Finally, we mention the $2$-outcome POVM with elements $M_{\pm} = \{\ket{\phi_{+}}\bra{\phi_{+}}, \ket{\phi_{-}}\bra{\phi_{-}} \}$, where
\begin{align}\label{eq:QCRBminMeNoS}
    \ket{\phi_{\pm}} &= (1\pm i)\bigg[ \left( \frac{1-C_0}{2} \mp \frac{D_0^{\Gamma}}{Q(\Gamma;\ket{\psi_{\Gamma}}\bra{\psi_{\Gamma}})^{1/2}} \right)\ket{a_0} \noindent\nonumber\\
    &~~~~~~ -\sum_{j>0}~\left(  \frac{C_j}{2} \pm \frac{D_j^{\Gamma}}{Q(\Gamma;\ket{\psi_{\Gamma}}\bra{\psi_{\Gamma}})^{1/2}}  \right) \ket{a_j}    \bigg], 
\end{align}
is the least susceptible to small measurement noise, in the sense of Ref.~\cite{kurdzialek2022measurement}. 
However, the projectors in Eq.~\eqref{eq:QCRBminMeNoS} are composed using modal amplitudes $C_j$ and derivatives $D_j^{\Gamma}$ that depends on the true value of $\Gamma$, which then necessitates the two-step adaptive estimation scheme previously outlined, as well as a more complicated pulse gating setup where the pump pulse must be shaped to match the projectors $\ket{\phi_{\pm}}$.
In contrast, mode-resolved photon counting $M_J$ represents a simpler~(if potentially more noise-prone) measurement strategy. 

\subsection{\texorpdfstring{$\langle \psi_{\Gamma} | \partial_{\Gamma} \psi_{\Gamma} \rangle \neq 0$}{}}

For $\langle \psi_{\theta} | \partial_{\theta} \psi_{\theta} \rangle \neq 0$~(see, for example, quadratically modulated pulses in Fig. \ref{fig:1cnew}), the fixed basis $M$ is no longer optimal as the corresponding $\Gamma$-SLD $L_{\Gamma}$ is no longer diagonal in the basis $M$.

In general, the corresponding optimal basis can be extracted by diagonalizing $L_{\Gamma}$, whose eigenvectors, and hence the optimal projectors, will depend on the value of the parameter $\Gamma$ itself. Just as for $M_{\pm}$ above, this corresponds to a more complicated measurement setup that requires a two-step adaptive estimation. 

\subsection{Results}

\begin{figure}[ht!]
        \includegraphics[width=0.4\textwidth]{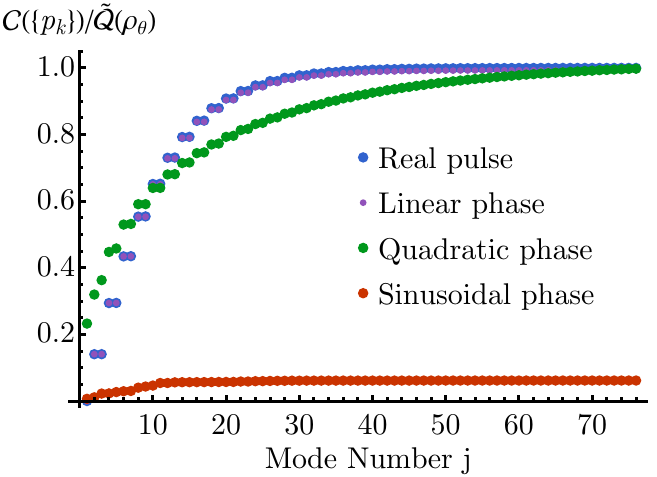}
        \caption{Ratio of cumulative mode-resolved photon counting Fisher information in Hermite-Gauss modes $\mathcal{C}(\{p_j\})$, with $\Gamma$-QFI $\mathcal{Q}(\rho_{\theta})$, for incoming unmodulated~(blue), linearly modulated~(purple), quadratically-modulated~(green), and sinusoidally modulated~(red) single-photon Fock state pulses with Gaussian envelope.  Parameter values set to $\Gamma T = 2.5$, $\Gamma_{\perp} = 5\Gamma, k= 1, \Omega =1$.}
        \label{fig:opt_measure1}
\end{figure}

Fig.~\ref{fig:opt_measure1} shows the predicted convergence~(following Eq.~(\ref{eq:predictedconverge})) of the CFI corresponding to measurement basis $M_{J}$
for unmodulated Gaussian pulses~(in blue), when $M_J$ is composed of Hermite-Gauss (HG) mode-resolved measurements. We also plot, on the same figure, the optimality ratio $\mathcal{C}(\{p_k\})/\mathcal{Q}(\rho_{\theta})$ for quadratically~(green) and sinusoidally-modulated~(red) incoming pulses, measured in the HG basis. While the HG basis is no longer optimal for either of the modulated pulses, the HG-mode-resolved CFI-to-QFI ratio is almost equal to unity for quadratic modulations for large enough mode number, owing to the fact the quadratically modulated pulses are still Gaussian. The sub-optimality is more obvious for a sinusoidally modulated pulse, where the CFI-to-QFI ratio is saturated to about $0.06$. Although it is possible to construct the complete set of $\mathcal{L}^2$ functions corresponding to quadratic and sinusoidal modulations of the incoming pulse, we see that the much more practicable HG basis still fetches a small but relatively significant proportion of the possible information about the $\Gamma$ parameter. 

Fig.~\ref{fig:opt_measure1} also demonstrates the sub-optimality of measurements in the fixed $M_J$, for the exemplary case of linear temporal modulation~(in purple) on real Gaussian pulses~(discussed in Sec. \ref{sec:lineaphase1}) where, although mode-resolved measurements in the $M_J$ basis are seen to achieve almost all of the QFI, the CFI-to-QFI ratio does not converge to unity. We also note in Fig.~\ref{fig:opt_measure1} that the origin of the sub-optimality of linearly modulated pulses is distinct from the sub-optimality associated with sinusoidally modulated pulses, both measured in the HG-mode-resolved basis -- the former is sub-optimal because there is \textit{no} fixed optimal basis corresponding to the pulse shape, owing to $\langle \psi_{\Gamma} | \partial_{\Gamma} \psi_{\Gamma} \rangle \neq 0$, whereas the latter is sub-optimal owing to the specific choice of measuring in the HG basis, which is only optimal for real pulses. An optimal fixed measurement basis for sinusoidally modulated pulses may be constructed, although they are expected to be a much less practical choice than the HG basis~\cite{eckstein2011quantum,donohue2018quantum,de2021effects,ansari2021achieving,garikapati2022programmable}.

Finally, we remark on the curious fact that the ratio $\mathcal{C}(\{p_k\})/\mathcal{Q}(\rho_{\theta})$ achieved using $M_{J}$ increases as the strength of detuning $|\Delta|$ increases~(not shown in the figure), meaning that even for incoming pulses with asymmetric spectral profiles, there may be scenarios in which the fixed basis $M_J$~(that would require significantly fewer resources than any measurement in parameter-dependent bases) are very close to optimal.

\section{Conclusions} \label{sec:Conclusions}

Temporal phase modulation introduces a rich variety of possibilities in single-photon pulse spectroscopy for estimating the coupling strength of TLS-photon interaction.
These are examined for different coupling geometries, represented by different values of the parameter $\Gamma_\perp$ relative to $\Gamma$, and also for distinct spectroscopies associated with different detection times.
For the class of Gaussian real pulses with quadratic phase modulation, in the long-time limit, the information obtained from both photon counting and wave-packet distortion depends on the incident pulse only through the spectral bandwidth.
This does not hold for general temporal phase and magnitude profiles.
When the TLS is partially excited, the diversity is more pronounced even for Gaussian pulses and an analysis motivated by experimental scenarios is advisable.

\begin{acknowledgments}
We thank Sourav Das for fruitful discussions and feedback on the manuscript. This work was supported, in part, by an EPSRC New Horizons grant (EP/V04818X/1) and the UKRI (Reference Number: 10038209) under the UK Government’s Horizon Europe Guarantee for the European Union's Horizon Europe Research and Innovation Programme under agreement 101070700 (MIRAQLS).
AK was supported by the Chancellor's International Scholarship from the University of Warwick.
FA acknowledges support from Marie Skłodowska-Curie Action EUHORIZON-MSCA-2021PF-01 (project QECANM, grant n. 701154).
Computing facilities were provided by the Scientific Computing Research Technology Platform of the University of Warwick.
\end{acknowledgments}

\bibliographystyle{apsrev4-2}
\bibliography{Refs}
 
\clearpage
\onecolumngrid
\appendix
\section{Single-photon QFI for complex-valued wave-packets: Time domain}\label{App:A} 

Finite-time QFI expression for $\xi(t) \in \mathbb{C}$ with regard to the normalized single-photon state is

\begin{align}
    \mathcal{Q}(\left| {{\rho }_{\Gamma }} \right\rangle )=\frac{{{\left( {{\partial }_{\Gamma }}{{p}_{\Gamma }} \right)}^{2}}}{{{p}_{\Gamma }}\left( 1-{{p}_{\Gamma }} \right)}+\left( 1-{{p}_{\Gamma }} \right)4\left( \left\langle  {{\partial }_{\Gamma }}{{\psi }_{\Gamma }} | {{\partial }_{\Gamma }}{{\psi }_{\Gamma }} \right\rangle -{{\left| \left\langle  {{\partial }_{\Gamma }}{{\psi }_{\Gamma }} | {{\psi }_{\Gamma }} \right\rangle  \right|}^{2}} \right),
\end{align}

where we have the normalized state  

\begin{align}
    \left| {{\psi }_{\Gamma }} \right\rangle =\frac{\left| {{\tilde{\psi} }_{g}^{P}}(t) \right\rangle }{\sqrt{\left\langle {{\tilde{\psi} }_{g}^{P}}(t)|{{\tilde{\psi} }_{g}^{P}}(t) \right\rangle }}\,\,\,\,\,\,,\,\,\,\,\,\,\,\,\left\langle  {{\psi }_{\Gamma }} | {{\psi }_{\Gamma }} \right\rangle =1 .
\end{align}

We can rewrite the QFI in terms of the unnormalized state of
\begin{align}
    \left| {{{\tilde{\psi }}}_{\Gamma }} \right\rangle =\sqrt{1-{{p}_{\Gamma }}}\left| {{\psi }_{\Gamma }} \right\rangle \,\,\,\,\,\,,\,\,\,\,\,\,\,\,\,\,\,\,\left\langle  {{{\tilde{\psi }}}_{\Gamma }} | {{{\tilde{\psi }}}_{\Gamma }} \right\rangle =1-{{p}_{\Gamma }},
\end{align}

and we obtain

\begin{align}\label{A4}
    \mathcal{Q}(\left| {{\rho }_{\Gamma }} \right\rangle )=\frac{{{\left( {{\partial }_{\Gamma }}{{p}_{\Gamma }} \right)}^{2}}}{{{p}_{\Gamma }}\left( 1-{{p}_{\Gamma }} \right)}\,+4\left\langle  {{\partial }_{\Gamma }}{{{\tilde{\psi }}}_{\Gamma }} | {{\partial }_{\Gamma }}{{{\tilde{\psi }}}_{\Gamma }} \right\rangle -\frac{4}{\left( 1-{{p}_{\Gamma }} \right)}{{\left| \left\langle  {{\partial }_{\Gamma }}{{{\tilde{\psi }}}_{\Gamma }} | {{{\tilde{\psi }}}_{\Gamma }} \right\rangle  \right|}^{2}} \equiv ~ \mathcal{C}(p_\Gamma) + \tilde{\mathcal{Q}}(\ket{\psi_\Gamma}).
\end{align}

The terms presented in \ref{A4} can be explicitly evaluated as follows:

\begin{align}
  & \left\langle  {{\partial }_{\Gamma }}{{{\tilde{\psi }}}_{\Gamma }} | {{\partial }_{\Gamma }}{{{\tilde{\psi }}}_{\Gamma }} \right\rangle =\int_{-\infty }^{t}{d}\tau \bigg[ \left( \int_{-\infty }^{\tau }{d}{t}'{{e}^{-\frac{\Gamma +{\Gamma_\perp}}{2}(\tau -{t}')}}{{\xi }^{*}}({t}') \right)\left( \int_{-\infty }^{\tau }{d}{t}'{{e}^{-\frac{\Gamma +{\Gamma_\perp}}{2}(\tau -{t}')}}\xi ({t}') \right) \\ \nonumber
 & -2\Gamma \operatorname{Re}\left[ \left( \int_{-\infty }^{\tau }{d}{t}'{{e}^{-\frac{\Gamma +{\Gamma_\perp}}{2}(\tau -{t}')}}{{\xi }^{*}}({t}') \right)\left( \int_{-\infty }^{\tau }{d}{t}'\frac{\tau -{t}'}{2}{{e}^{-\frac{\Gamma +{\Gamma_\perp}}{2}(\tau -{t}')}}\xi ({t}') \right) \right] \\ \nonumber
 &+{{\Gamma }^{2}}\left( \int_{-\infty }^{\tau }{d}{t}'\frac{\tau -{t}'}{2}{{e}^{-\frac{\Gamma +{\Gamma_\perp}}{2}(\tau -{t}')}}{{\xi }^{*}}({t}') \right)\left( \int_{-\infty }^{\tau }{d}{t}'\frac{\tau -{t}'}{2}{{e}^{-\frac{\Gamma +{\Gamma_\perp}}{2}(t-{t}')}}\xi ({t}') \right)\bigg],  
\end{align}

\begin{align}
    \left\langle  {{\partial }_{\Gamma }}{{{\tilde{\psi }}}_{\Gamma }} | {{{\tilde{\psi }}}_{\Gamma }} \right\rangle =\int_{-\infty }^{t}{d}\tau \bigg[& \left( \xi (\tau )-\Gamma \left( \int_{-\infty }^{\tau }{d}{t}'{{e}^{-\frac{\Gamma +{\Gamma_\perp}}{2}(\tau -{t}')}}\xi ({t}') \right) \right) \\ \nonumber
    &\left( \int_{-\infty }^{\tau }{d}{t}'\left( \frac{\Gamma \tau -\Gamma {t}'}{2}-1 \right){{e}^{-\frac{\Gamma +{\Gamma_\perp}}{2}(\tau -{t}')}}{{\xi }^{*}}({t}') \right) \bigg],
\end{align}

\begin{align}
{{p}_{\Gamma }}={{\left| {{\psi }_{e}(t)} \right|}^{2}}+\left\langle {{\tilde{\psi} }_{g}^{E}}(t)|{{\tilde{\psi} }_{g}^{E}}(t) \right\rangle,
\end{align}

\begin{align}
    {{\partial }_{\Gamma }}{{p}_{\Gamma }}=\psi _{e}^{*}{{\partial }_{\Gamma }}{{\psi }_{e}}+{{\psi }_{e}}{{\partial }_{\Gamma }}\psi _{e}^{*}+\left\langle {{\partial }_{\Gamma }}{{\tilde{\psi} }_{g}^{E}}(t)|{{\tilde{\psi} }_{g}^{E}}(t) \right\rangle +\left\langle {{\tilde{\psi} }_{g}^{E}}(t)|{{\partial }_{\Gamma }}{{\tilde{\psi} }_{g}^{E}}(t) \right\rangle,
\end{align}

\begin{align}
    {{\partial }_{\Gamma }}{{p}_{\Gamma }}=\psi _{e}^{*}(t){{\partial }_{\Gamma }}{{\psi }_{e}}(t)+{{\psi }_{e}}(t){{\partial }_{\Gamma }}\psi _{e}^{*}(t)+{\Gamma_\perp}\int_{-\infty }^{t}{d}\tau \,{{\psi }_{e}}(\tau ){{\partial }_{\Gamma }}\psi _{e}^{*}(\tau )+{\Gamma_\perp}\int_{-\infty }^{t}{d}\tau \,\psi _{e}^{*}(\tau ){{\partial }_{\Gamma }}{{\psi }_{e}}(\tau),
\end{align}

\begin{align}
  & {{\partial }_{\Gamma }}{{p}_{\Gamma }}\left( t \right)=-\left( \int_{-\infty }^{t}{d}{t}'{{e}^{-\frac{\Gamma +{\Gamma_\perp}}{2}(t-{t}')}}{{\xi }^{*}}({t}') \right)\left( \int_{-\infty }^{t}{d}{t}'\frac{\Gamma t-\Gamma {t}'-1}{2}{{e}^{-\frac{\Gamma +{\Gamma_\perp}}{2}(t-{t}')}}\xi ({t}') \right) \\ \nonumber
 &-\left( \int_{-\infty }^{t}{d}{t}'{{e}^{-\frac{\Gamma +{\Gamma_\perp}}{2}(t-{t}')}}\xi ({t}') \right)\left( \int_{-\infty }^{t}{d}{t}'\frac{\Gamma t-\Gamma {t}'-1}{2}{{e}^{-\frac{\Gamma +{\Gamma_\perp}}{2}(t-{t}')}}{{\xi }^{*}}({t}') \right) \\ \nonumber
 & +{\Gamma_\perp}\int_{-\infty }^{t}{d}\tau \bigg(-\left( \int_{-\infty }^{\tau }{d}{t}'{{e}^{-\frac{\Gamma +{\Gamma_\perp}}{2}(\tau -{t}')}}{{\xi }^{*}}({t}') \right)\left( \int_{-\infty }^{\tau }{d}{t}'\frac{\Gamma \tau -\Gamma {t}'-1}{2}{{e}^{-\frac{\Gamma +{\Gamma_\perp}}{2}(\tau -{t}')}}\xi ({t}') \right) \\ \nonumber
 &-\left( \int_{-\infty }^{\tau }{d}{t}'{{e}^{-\frac{\Gamma +{\Gamma_\perp}}{2}(\tau -{t}')}}\xi ({t}') \right)\left( \int_{-\infty }^{\tau }{d}{t}'\frac{\Gamma \tau -\Gamma {t}'-1}{2}{{e}^{-\frac{\Gamma +{\Gamma_\perp}}{2}(\tau -{t}')}}{{\xi }^{*}}({t}') \right)\bigg).
\end{align}
\section{Input Pulse Shapes for which \texorpdfstring{$\langle \psi_{\Gamma} | \partial_{\Gamma} \psi_{\Gamma} \rangle = 0$}{}}\label{app:realvec_pulses}

The general class of input functions $\tilde{\xi}(\omega)$ for which the overlap $\langle \psi_{\theta} | \partial_{\Gamma} \psi_{\Gamma} \rangle = 0$, related to the second part of the QFI for the outgoing modified single-photon wavepacket in Eq.~(\ref{eq:qfipure}), can be obtained explicitly for TLS estimation. Starting with the general form of the normalized, modified single-photon wavepacket, $\ket{\psi_{\Gamma}} = \frac{1}{\sqrt{1-p_{\Gamma}}}\,\ket{\tilde{\psi}_{\Gamma}}$, we obtain the parametric derivative as 
\begin{equation}
    \ket{\partial_{\Gamma}\psi_{\Gamma}} = \frac{1}{\sqrt{1-p_{\Gamma}}}\,\ket{\partial_{\Gamma}\tilde{\psi}_{\Gamma}} + \frac{\partial_{\Gamma}p_{\Gamma}}{2(1-p_{\Gamma})^{3/2}}\,\ket{\tilde{\psi}_{\Gamma}}
\end{equation}
The overlap is then
\begin{equation}
    \langle \psi_{\Gamma}|\partial_{\Gamma}\psi_{\Gamma}\rangle = \frac{1}{1-p_{\Gamma}}\,\langle \tilde{\psi}_{\Gamma}|\partial_{\Gamma}\tilde{\psi}_{\Gamma}\rangle + \frac{\partial_{\Gamma}p_{\Gamma}}{2(1-p_{\Gamma})}
\end{equation}
From Eq.~(\ref{eq:qfigamma2}), we have explicitly for $\Gamma$-estimation for TLS:
\begin{equation}\label{eq:overlapunnorm}
    \langle \tilde{\psi}_{\Gamma}|\partial_{\Gamma}\tilde{\psi}_{\Gamma}\rangle = \int d\omega |\tilde{\xi}(\omega)|^2\,\,\left[  
    \frac{\Gamma_{\perp}}{2}\left(  \frac{\frac{\Gamma^2-\Gamma_{\perp}^2}{4} - \omega^2}{ \left( \frac{(\Gamma+\Gamma)^2)}{4} + \omega^2\right)^2  } + i\omega \frac{ \frac{\Gamma^2+\Gamma_{\perp}^2}{4} + \omega^2 } { \left( \frac{(\Gamma+\Gamma)^2)}{4} + \omega^2\right)^2}   \right)       \right]
\end{equation}
For $|\tilde{\xi}(\omega)|^2 = |\tilde{\xi}(-\omega)|^2$, the overlap $\langle \tilde{\psi}_{\Gamma}|\partial_{\Gamma}\tilde{\psi}_{\Gamma}\rangle$ in Eq.~(\ref{eq:overlapunnorm}) is then necessarily real, implying then that $\langle \psi_{\Gamma}|\partial_{\Gamma}\psi_{\Gamma}\rangle \in \mathds{R}$. However, derivating the normalization condition $\langle \psi_{\Gamma}|\psi_{\Gamma}\rangle = 1$ with respect to $\Gamma$, we see that $\mathrm{Re} \langle \psi_{\Gamma}|\partial_{\Gamma}\psi_{\Gamma}\rangle = 0$, thus showing that for all incoming pulses whose magnitudes are symmetric in the frequency domain, $\langle \psi_{\Gamma}|\partial_{\Gamma}\psi_{\Gamma}\rangle = 0$. 

This family then includes pulses with real amplitudes in the time domain~(whose frequency domain amplitude then obey $\tilde{\xi}(\omega) = \tilde{\xi}^{*}(-\omega)$, but also certain classes of temporally modulated pulses considered in this paper -- both quadratic and sinusoidal modulations~(see Fig. \ref{fig:1anew}) preserve the symmetry of the $|\tilde{\xi}(\omega)|^2$, meaning that the overlap for these modulations vanishes. 

On the other hand, for the exponential pulse that has been quadratically modulated~(see Fig. \ref{fig:1cnew}), we see that the resulting spectral distribution is no longer symmetric in $\omega$, implying that  $\langle \psi_{\Gamma} | \partial_{\Gamma} \psi_{\Gamma} \rangle \neq 0$. Finally, we also note that a linear detuning would cause a shift as $\omega \rightarrow \omega - \Delta$ in the integrand of Eq.~(\ref{eq:overlapunnorm}). This implies that for the general class of detuned, linearly shifted pulses, we have $\langle \psi_{\Gamma} | \partial_{\Gamma} \psi_{\Gamma} \rangle \neq 0$. 

\section{TLS-pulse-environment states for complex-valued wave-packets: Frequency domain}\label{App:B} 

The excitation amplitude of the TLS, unnormalized single-photon states in the pulse and environment modes, can be expressed as:

\begin{align}
{{\psi }_{e}}(t)=-\frac{1}{\sqrt{2\pi}}\int_{-\infty }^{\infty }{d}\omega  {{e}^{-\I\omega t}} \tilde{\xi} (\omega) f(\omega),
\end{align}
\begin{align}
\left| {{\psi }_{g}^{P}}(t) \right\rangle =\int_{-\infty }^{\infty }{d}\omega {{\psi }_{g}^{P}}(t,\omega ){{a}^{\dagger }}(\omega )\left| {{0}^{P}} \right\rangle, 
\end{align}
\begin{align}
\left| {{\psi }_{g}^{E}}(t) \right\rangle =\int_{-\infty }^{\infty }{d}\omega {{\psi }_{g}^{E}}(t,\omega ){{b}^{\dagger }}(\omega )\left| {{0}^{E}} \right\rangle, 
\end{align}
where,
\begin{align}
{{\psi }_{g}^{P}}(t,\omega )=\tilde{\xi} (\omega )-\frac{1}{2}\sqrt{\Gamma} \tilde{\xi} (\omega) f(\omega) +\frac{\I\sqrt{\Gamma} }{2\pi }\int_{-\infty }^{\infty }{d{\omega }'} \tilde{\xi}(\omega') f(\omega') \frac{{{e}^{\I\left( \omega -{\omega }' \right)t}}}{\omega -{\omega }'},
\end{align}
\begin{align}
{{\psi }_{g}^{E}}(t,\omega )=-\frac{\sqrt{{{\Gamma }_{\perp }}}}{2}\tilde{\xi} (\omega) f(\omega) +\frac{\I\sqrt{{{\Gamma }_{\perp }}}}{2\pi }\int_{-\infty }^{\infty }{d{\omega }'} \tilde{\xi}(\omega') f(\omega') \frac{{{e}^{\I\left( \omega -{\omega }' \right)t}}}{\omega -{\omega }'}.
\end{align}
If $t\to \infty $, then we have,
\begin{align}
{{\psi }_{e}}(\infty )=0,
\end{align}
\begin{align}
{{\psi }_{g}^{P}}(\infty ,\omega )=\tilde{\xi} (\omega )-\sqrt{\Gamma} \tilde{\xi}(\omega) f(\omega) ,
\end{align}
\begin{align}
{{\psi }_{g}^{E}}(\infty ,\omega )=-\sqrt{\Gamma_\perp} \tilde{\xi}(\omega) f(\omega).
\end{align}

\section{Analytical expressions of the asymptotic QFI for a Gaussian pulse and its quadratically phased counterpart} \label{R vs PM}

CFI for a real Gaussian pulse, denoted by $C_R$, can be determined using Eqs. (\ref{eq:QFI_1ph_normalized}), (\ref{eq:pgamma}), and (\ref{eq:dpgamma}). It can be expressed as:

\begin{align}\label{eq:CFI_gau_A}
    \Gamma^2\mathcal{C}_R\left({\rho_\Gamma} \right)=-\frac{\gamma e^{-\frac{(\gamma+1)^2}{8 \sigma_\omega ^{2}}} \left(\sqrt{2 \pi } e^{\frac{(\gamma+1)^2}{8 \sigma_\omega ^{2}}} \left(4 \gamma \sigma_\omega ^{2}+(\gamma+1)^2\right) \text{erfc}\left(\frac{\gamma+1}{2 \sqrt{2} \sigma_\omega }\right)-4 (\gamma+1) \sigma_\omega \right)^2}{16 \sqrt{2 \pi } (\gamma+1)^2 \sigma_\omega ^{4} \text{erfc}\left(\frac{\gamma+1}{2 \sqrt{2} \sigma_\omega }\right) \left(\sqrt{2 \pi } \gamma e^{\frac{(\gamma+1)^2}{8 \sigma_\omega ^{2}}} \text{erfc}\left(\frac{\gamma+1}{2 \sqrt{2} \sigma_\omega }\right)-(\gamma+1) \sigma_\omega \right)},
\end{align}
where $\gamma=\Gamma_\perp/\Gamma$ and $\sigma_\omega =(1/(2\Gamma T))$.

The quantum contribution for a real Gaussian pulse denoted as $\tilde{Q}_R$, is determined using Eqs. (\ref{eq:qfipure}), (\ref{eq:qfigamma1}), and (\ref{eq:qfigamma2}). It can be expressed as follows: 

\begin{align}\label{eq:QFI_gau_A}
    &\Gamma^2\tilde{\mathcal{Q}}_R\left({\ket{\psi_\Gamma}} \right)=\frac{1}{16 (\gamma+1)^3 \sigma_\omega ^{5}}\Bigg( -\frac{\gamma^2 \left(\sqrt{2 \pi } e^{\frac{(\gamma+1)^2}{8 \sigma_\omega ^{2}}} \left(4 \gamma \sigma_\omega ^{2}+(\gamma+1)^2\right) \text{erfc}\left(\frac{\gamma+1}{2 \sqrt{2} \sigma_\omega }\right)-4 (\gamma+1) \sigma_\omega \right)^2}{\gamma \sigma_\omega -\sqrt{2 \pi } \gamma e^{\frac{(\gamma+1)^2}{8 \sigma_\omega ^{2}}} \text{erfc}\left(\frac{\gamma+1}{2 \sqrt{2} \sigma_\omega }\right)+\sigma_\omega } \\ \nonumber
    &+8 \sigma_\omega ^{2} \left(\sqrt{2 \pi } e^{\frac{(\gamma+1)^2}{8 \sigma_\omega ^{2}}} \left(4 (2 \gamma (\gamma+1)+1) \sigma_\omega ^{2}+(2 \gamma+1) (\gamma+1)^2\right) \text{erfc}\left(\frac{\gamma+1}{2 \sqrt{2} \sigma_\omega }\right)-4 (\gamma+1) (2 \gamma+1) \sigma_\omega \right) \Bigg).
\end{align}

Also, the probability of a single photon surviving, denoted as $p_\Gamma$, for a real Gaussian pulse can be expressed as:
\begin{align}
    p_\Gamma = \frac{\sqrt{2 \pi } \gamma e^{\frac{(\gamma+1)^2}{8 \sigma_\omega ^{2}}} \text{erfc}\left(\frac{\gamma+1}{2 \sqrt{2} \sigma_\omega }\right)}{(\gamma+1) \sigma_\omega}.
\end{align}

One can obtain the analytical expressions for the quadratically phase-modulated Gaussian pulse by substituting $\sigma_\omega$ with
$ \sigma' _{\omega }= \sqrt{1+16 k^2 \Gamma^4T^4}\sigma_\omega$.
\section{Analytical expressions of the asymptotic QFI for the linearly phased decaying exponential pulse}
\label{App:decaying}
CFI for the linearly phased decaying exponential pulse, using Eqs. (\ref{eq:QFI_1ph_normalized}),(\ref{eq:pgamma}), and (\ref{eq:dpgamma}), can be expressed as:
\begin{align}\label{CFI-Decay}
    \Gamma^2\mathcal{C}_R\left({\rho_\Gamma} \right)=\frac{\gamma \Gamma T \left(32 \Delta ^2 T^2 \left(\gamma+(\gamma+1)^2 T\right)+8 \left(\gamma+\left(\gamma^2-1\right) \Gamma T\right) (\gamma \Gamma T+\Gamma T+1)^2\right)^2}{16 (\gamma+1)^3 (\gamma \Gamma T+\Gamma T+1) \left(4 \Delta ^2 T^2+(\gamma \Gamma T+\Gamma T+1)^2\right)^3 \left(1-\frac{4 \gamma \Gamma T (\gamma \Gamma T+\Gamma T+1)}{(\gamma+1) \left(4 \Delta ^2 T^2+(\gamma \Gamma T+\Gamma T+1)^2\right)}\right)},
\end{align} 
The quantum contribution for the linearly phased decaying exponential pulse, using Eqs. (\ref{eq:qfipure}),(\ref{eq:qfigamma1}), and (\ref{eq:qfigamma2}),  can be expressed as:
\begin{align}\label{QFI-Decay}
    \Gamma^2\tilde{\mathcal{Q}}_R\left({\ket{\psi_\Gamma}} \right)=& 8 \Gamma T \bigg( 2 \gamma^3+4 \gamma^2+3 \gamma+(\gamma+1) \Gamma^2T^2 \left( 2 \gamma^4+2 \gamma^3+\gamma^2 \left(8 \Delta ^2/\Gamma^2 +3\right) +8 \gamma \Delta ^2/\Gamma^2 +4 \Delta ^2/\Gamma^2 +1\right)\\ \nonumber
    +&\left(4 \gamma^4+8 \gamma^3+8 \gamma^2+4 \gamma+2\right) \Gamma T+1 \bigg) \bigg/ \bigg( (\gamma+1)^3 \left(4 \Delta ^2 T^2+(\gamma \Gamma T+\Gamma T+1)^2\right) \\ \nonumber
    &\left(\gamma+(\gamma+1) \Gamma^2T^2 \left(\gamma^2-2 \gamma+4 \Delta^2/\Gamma^2 +1\right)+2 \left(\gamma^2+1\right) \Gamma T+1\right) \bigg).
\end{align} 
\section{Modal Decomposition}\label{app:modaldecom}
 An alternative form for $\mathcal{Q}(\ket{\psi_{\theta}})$ can be obtained by transforming to a set of discrete basis functions, which we will refer to as the modal picture henceforth. Inserting a complete set of basis modes in the frequency domain $ \sum_n g_n^*(\omega) g_n(\omega') = \delta(\omega-\omega')$ into the definition of the white noise operators in Eq.~(\ref{eq:white_noise_a}), we get
\begin{equation}
    a(t) = \sum_n g^*_n(t)a_n, 
\end{equation}
where 
\begin{equation}\label{eq:modalopdef}
    a_n = \int d\omega\, g_n(\omega)\, a(\omega)\,, ~~[a_m,a_n^{\dag}] = \delta_{mn}
\end{equation}
is the $n$-mode annihilation operator, and $g_n(t)$ is the Fourier transformed discrete basis function,
\begin{equation}
    g_n^*(t) = \frac{1}{\sqrt{2\pi}}~\int d\omega e^{-i\omega t}g_n^*(\omega).
\end{equation}
Although the choice of the complete basis is arbitrary, the algebra is much simplified if we pick a basis of which the incoming mode envelope function $\xi(t)$ is a member. This is possible, as one can always construct a complete, orthogonal basis starting from the envelope function using the Gram-Schmidt procedure for square-integrable functions on $L^2(\mathbb{R})$.  
 The incoming state is then simply assumed to be the state created by the zeroth-mode creation operator,
\begin{equation}\label{eq:onephotonmodaldef}
    \ket{1}_{\xi} = a_0^{\dag}\ket{0}, a^{\dag}_0 = \int d\omega\, g^*_0(\omega)\, a^{\dag}(\omega)\ = \int d\omega\, \tilde{\xi}(\omega)\, a^{\dag}(\omega).
\end{equation}
The outgoing post-selected one-photon state in terms of this modal decomposition is then given by
\begin{align}\label{eq:outgoingonephotonmode}
     \ket{\psi_{\theta}} &= \frac{1}{\sqrt{1-p_{\theta}}}\left[(1-C_0)\,a^{\dag}_0 - \sum_{j>0}~C_j\, a^{\dag}_j  \right]\ket{0^{\mathrm{P}}}
\end{align} 
where, in terms of the mode components, 
\begin{equation}\label{eq:gcoeffdef}
    C_j = \Gamma\,\int_{t_0}^{t} dt_1\int_{t_0}^{t_1}dt_2~\mathrm{exp}\left[-\left(\frac{\Gamma+\Gamma_{\perp}}{2} - i\Delta\right)(t_1-t_2)\right]~g_j(t_1)g_0^{*}(t_2)
\end{equation}
are modal amplitudes in the scattered state, obtained using the commutation relation for the modal operators in Eq.~(\ref{eq:modalopdef}). 
Abbreviating the $\theta$-derivatives as $D_j^{\theta} = \partial_{\theta}C_j$, the quantum contribution to the QFI of the outgoing state in the modal decomposition can also be obtained
\begin{align}
    (1-p_{\theta})\mathcal{Q}(\ket{\psi_{\theta}})= 
    4\sum_j\,|D_j^{\theta}|^2 - \frac{4}{1-p_{\theta}}\,\mathrm{Im}\left( (1-C_0^*)D_0^{\theta} - \sum_{j>0}\,C_j^* D_j^{\theta}    \right)^2 - \frac{(\partial_{\theta}p_{\theta})^2}{1-p_{\theta}} 
\end{align}
The outgoing QFI is then
\begin{align}\label{eq:Groundqfimodal}
    \mathcal{Q}(\rho_{\theta}) = 
    4\sum_j\,|D_j^{\theta}|^2 - \frac{4}{1-p_{\theta}}\,\mathrm{Im}\left( (1-C_0^*)D_0^{\theta} - \sum_{j>0}\,C_j^* D_j^{\theta}    \right)^2 + \frac{(\partial_{\theta}p_{\theta})^2}{p_{\theta}} .
\end{align}

\end{document}